\documentclass[aps,twocolumn,nofootinbib,groupedaddress,superscriptaddress,longbibliography]{revtex4-1}

\usepackage{amsmath,amssymb}
\usepackage[normalem]{ulem}
\usepackage{graphicx}
\usepackage{multirow,makecell}
\usepackage{bm}
\usepackage{mathtools}
\usepackage{dsfont}
\usepackage{amsfonts}
\usepackage{xcolor}
\usepackage{url}
\usepackage{rotating}
\usepackage{wasysym}
\usepackage{diagbox}

\usepackage[colorlinks=true,linkcolor=blue,citecolor=blue,urlcolor=blue,hypertexnames=false]{hyperref}
\newcommand{\nocontentsline}[3]{}
\newcommand{\tocless}[2]{\bgroup\let\addcontentsline=\nocontentsline#1{#2}\egroup}

\def\ba#1\ea{\begin{align}#1\end{align}}
\def\bg#1\eg{\begin{gather}#1\end{gather}}
\def\bpm{\begin{pmatrix}}
\def\epm{\end{pmatrix}}

\newcommand{\nn}{\nonumber \\ }
\newcommand{\td}[1]{\widetilde{#1}}

\newcommand{\hf}{\frac{1}{2}}

\newcommand{\ls}[1]{{}_{#1}}
\newcommand{\hfx}[1]{\frac{#1}{2}}

\newcommand{\bb}[1]{{\mathbf #1}}

\newcommand{\bk}{\bb k}
\newcommand{\bR}{\bb R}
\newcommand{\bQ}{\bb Q}
\newcommand{\bq}{\bb q}
\newcommand{\bba}{\bb a}

\newcommand{\bS}{\bb S}

\newcommand{\mc}[1]{\mathcal{#1}}
\newcommand{\mf}[1]{\mathfrak{#1}}
\newcommand{\der}{\partial}
\newcommand{\dg}{\dagger}
\newcommand{\om}{\omega}
\newcommand{\sg}{\sigma}

\newcommand{\vep}{\varepsilon}

\newcommand{\nc}{N_c}
\newcommand{\norb}{N_{orb}}
\newcommand{\nocc}{N_{occ}}
\newcommand{\ntot}{N_{tot}}
\newcommand{\al}{\alpha}
\newcommand{\be}{\beta}
\newcommand{\cdag}{c^{\dg}}
\newcommand{\ua}{\uparrow}
\newcommand{\da}{\downarrow}
\newcommand{\gs}{\ket{\Phi}}
\newcommand{\Emag}{\mc{E}(\bQ)}
\newcommand{\Heff}{\mc{H}^{SE}_{\bk \al n,\bk' \be n'}(\bQ)}
\newcommand{\Uc}{\frac{U}{\nc}}
\newcommand{\Hse}{\mc{H}^{SE}}
\newcommand{\Em}{\mc{E}_G(\bQ)}
\newcommand{\Eg}{\mc{E}_G(\bQ)}
\newcommand{\Dmag}{D_{\mu \nu}}
\newcommand{\Dpert}{D_{\mu \nu}^{(1)}}
\newcommand{\Epert}{\mc{E}_G^{(1)} (\bQ)}
\newcommand{\fid}{\chi_{\mu \nu}^{m n} (\bk)}
\newcommand{\qgt}{\mf{G}_{\mu \nu} (\bk)}
\newcommand{\qm}{\mf{g}_{\mu\nu}(\bk)}
\newcommand{\bcurv}{\mc{F}_{\mu\nu}(\bk)}

\newcommand{\ee}[1]{e^{#1}}
\newcommand{\uc}{\frac{U}{\nc}}
\newcommand{\smax}{S_{max}}
\newcommand{\pdag}{\psi^{\dg}}
\newcommand{\ketq}{\ket{\bQ}}
\newcommand{\braq}{\bra{\bQ}}
\newcommand{\ddmag}{D_{\mu \mu}}
\newcommand{\ddpert}{D^{(1)}_{\mu \mu}}

\newcommand{\limu}{\lim_{U \rightarrow \infty}}

\newcommand{\ket}[1]{|#1\rangle}
\newcommand{\bra}[1]{\langle#1|}
\newcommand{\brk}[2]{\langle#1|#2\rangle}
\newcommand{\kbr}[2]{|#1\rangle\langle#2|}
\newcommand{\normsq}[1]{\langle#1|#1\rangle}
\newcommand{\expv}[1]{\langle#1\rangle}

\newcommand{\brkr}[2]{\langle#1|#2\rangle}

\newcommand{\brao}{\bra{\Phi}}
\newcommand{\keto}{\ket{\Phi}}
\newcommand{\Pdag}{\Psi_{\bQ}^{\dg}}

\newcommand{\zst}{z^*_{\bk \al n}(\bQ)}

\newcommand{\den}[1]{\frac{1}{#1}}
\newcommand{\dm}{\der_\mu}
\newcommand{\dn}{\der_\nu}
\newcommand{\um}{u_m(\bk)}
\newcommand{\un}{u_n(\bk)}
\newcommand{\emm}{E_m(\bk)}
\newcommand{\qmn}{\frac{Q_\mu Q_\nu}{2 \ntot}}
\newcommand{\sumk}{\sum_{\bk}^{\nc}}
\newcommand{\summ}{\sum_m^{\norb}}
\newcommand{\sumnn}{\sum_n^{\nocc}}

\newcommand{\enn}{E_n(\bk)}
\newcommand{\mnarrow}{\mu \leftrightarrow \nu}

\newcommand{\vac}{\ket{0}}
\newcommand{\uk}{\mc{U}(\bk)}
\newcommand{\ukdag}{\uk^{\dg}}
\newcommand{\id}{\mathds{1}}
\newcommand{\proj}{P(\bk)}
\newcommand{\ehat}{\hat{\bb e}}
\newcommand{\qgtmat}{\mf{G}(\bk)}
\newcommand{\gmat}{\mf{g}(\bk)}
\newcommand{\fmat}{\mc{F}(\bk)}
\newcommand{\qdist}{s(\bk,\bk')^2}
\newcommand{\rar}{\Rightarrow}

\newcommand{\lrar}{\Leftrightarrow}
\newcommand{\psd}{\succeq}
\newcommand{\pdef}{\succ}

\newcommand{\Egb}{\mc{E}_G}
\newcommand{\Etd}{\td{E}}
\newcommand{\pcurv}{\mc{P}}
\newcommand{\Ptd}{\td{P}}
\newcommand{\lmax}{l_{max}}
\newcommand{\ltot}{l_{tot}}

\newcommand{\Dsec}{\Dmag^{(2)}}
\newcommand{\Esec}{\mc{E}_G^{(2)}(\bQ)}

\newcommand{\eq}[1]{Eq.~\eqref{#1}}

\newcommand{\fig}[1]{Fig.~\ref{#1}}

\newcommand{\ourtitle}{
Quantum geometric bound for saturated ferromagnetism
}

\allowdisplaybreaks

\begin{document}
\title{\textbf{\ourtitle}}

\author{Junha \surname{Kang}}
\affiliation{Department of Physics and Astronomy, Seoul National University, Seoul 08826, Korea}
\affiliation{Center for Theoretical Physics (CTP), Seoul National University, Seoul 08826, Korea}
\affiliation{Institute of Applied Physics, Seoul National University, Seoul 08826, Korea}

\author{Taekoo \surname{Oh}}
\affiliation{RIKEN Center for Emergent Matter Science (CEMS), Wako, Saitama 351-0198, Japan}

\author{Junhyun \surname{Lee}}
\affiliation{Department of Physics and Astronomy, Center for Materials Theory,
	Rutgers University, Piscataway, NJ 08854, United States of America}

\author{Bohm-Jung \surname{Yang}}
\email{bjyang@snu.ac.kr}
\affiliation{Department of Physics and Astronomy, Seoul National University, Seoul 08826, Korea}
\affiliation{Center for Theoretical Physics (CTP), Seoul National University, Seoul 08826, Korea}
\affiliation{Institute of Applied Physics, Seoul National University, Seoul 08826, Korea}

\begin{abstract}
Despite its abundance in nature, predicting the occurrence of ferromagnetism in the ground state is possible only under very limited conditions such as in a flat band system with repulsive interaction or in a band with a single hole under infinitely large Coulomb repulsion, etc.
Here, we propose a general condition to achieve saturated ferromagnetism based on the quantum geometry of electronic wave functions in itinerant electron systems.
By analyzing the spin excitations of multi-band repulsive Hubbard models with an integer band filling, relevant to either ferromagnetic insulators or semimetals, we show that quantum geometry stabilizes the Goldstone mode in the strongly correlated limit.
Our theory indicates the stability of ferromagnetism in a large class of insulators and semimetals other than the previously studied flat band systems and their variants.
Moreover, we rigorously prove that saturated ferromagnetism is forbidden in any system with trivial quantum geometry, which includes every half-filled system.
We believe that our findings reveal a profound connection between quantum geometry and ferromagnetism, which can be extended to various symmetry-broken ground states in itinerant electronic systems.
\end{abstract}

\maketitle

\let\oldaddcontentsline\addcontentsline
\renewcommand{\addcontentsline}[3]{}


\textit{Introduction.---}
Ferromagnetism is the simplest form of symmetry-broken ground states whose fundamental origin has been discussed by divergent perspectives, from Heisenberg's localized electron picture~\cite{Heisenberg1928} to Bloch's itinerant electron approach~\cite{bloch1929bemerkung}.
More recently, the modern theory of ferromagnetism formulated based on the Hubbard model~\cite{gutzwiller1963effect,kanamori1963electron,hubbard1963electron} has led to various exact theorems including the following two distinct mechanisms.
One is Nagaoka's ferromagnetism~\cite{nagaoka1966ferromagnetism,tasaki1989extension,tasaki1998nagaoka}, where a single hole in the valence band induces the spin alignment when the electron repulsion is infinite.
Although mathematically interesting, Nagaoka's ferromagnetism is considered to be too fragile to exist in realistic electronic systems, as it is destroyed upon introducing a second hole~\cite{shastry1990instability}.
The contemporary understanding is largely based on flat band ferromagnetism and its variants extensively studied by Mielke~\cite{mielke1991ferromagnetic,mielke1992exact,mielke1993ferromagnetism,mielke1999stability} and Tasaki~\cite{tasaki1992ferromagnetism,mielke1993ferromagnetism,tasaki1995ferromagnetism,tasaki2003ferromagnetism,tasaki1994stability,tasaki1996stability,tanaka2020extension}, where the authors construct exactly solvable models that rigorously exhibit ferromagnetism.
Despite delivering valuable insights into the origin of ferromagnetism, the requirement of a flat band places a heavy constraint regarding its applicability to general systems.
Thus, it is imperative to obtain a general criterion for the stability of ferromagnetism applicable to the vast majority of systems where such theorems are out of reach.
Recently, the quantum geometry of electronic wave functions has garnered increasing attention due to its relevance to various physical phenomena,
such as anomalous Landau levels of flat bands \cite{rhim2020quantum,hwang2021geometric,jung2024quantum}, flat band superconductivity \cite{liang2017band,xie2020topology,tian2023evidence}, and nonlinear Hall effect \cite{gao2023quantum,wang2023quantum}, etc.
Moreover, it has been shown that quantum geometry plays a prominent role in describing the interacting ground states with spontaneous symmetry breaking~\cite{herzog2022many,han2024quantum}.
For instance, in superconductors, quantum geometry was shown to be deeply related to the superfluid weight~\cite{peotta2015superfluidity,xie2020topology}, the length scale of Cooper pairs \cite{hu2023anomalous,chen2024ginzburg}, and the dynamics of the Higgs mode~\cite{villegas2021anomalous}.
Also, in excitonic ground states, quantum geometry was shown to induce anomalous Lamb shifts \cite{srivastava2015signatures}, contribute to the exciton drift velocity \cite{cao2021quantum}, and stabilize exciton condensates \cite{hu2022quantum}.
In contrast, the role of quantum geometry in magnetic ground states remains largely unexplored~\cite{bernevig2021twisted,wu2020quantum,torma2023essay,herzog2022many}.
In this Letter, we construct an exact criterion for achieving saturated ferromagnetism (SFM) in repulsive Hubbard models, applicable to any integer-filled kinetic Hamiltonian with arbitrary dimensionality and complexity, given that it becomes an insulator or semimetal upon filling one spin sector.
By examining the spin excitations of the Hubbard model, we obtain an analytic expression for the Goldstone mode valid in the strongly correlated limit.
We show that the stability of the Goldstone mode is governed by the quantum geometry of the non-interacting electron bands, and express the spin stiffness, which is the inverse mass of the Goldstone boson, in terms of the band-resolved quantum geometric tensors~\cite{watanabe2021chiral,kitamura2022quantum}.
This shows that the ferromagnetism of insulators or semimetals arises from the delicate interplay between Wannier function spreading and the Coulomb repulsion, similar to Heisenberg's ferromagnetism~\cite{Heisenberg1928}.
Moreover, by identifying the aforementioned expression as a rigorous upper bound of the Goldstone mode, we show that saturated ferromagnetism is strictly prohibited when the quantum geometry is trivial, which formally rules out the possibility of saturated ferromagnetism in every half-filled Hubbard model.

\textit{Saturated ferromagnetism.---}
SFM represents the most robust manifestation of ferromagnetic behavior, characterized by the ground state possessing maximum total spin~\cite{nagaoka1966ferromagnetism,tasaki1989extension,tasaki1998nagaoka,mielke1991ferromagnetic,mielke1992exact,mielke1993ferromagnetism,mielke1999stability,tasaki1992ferromagnetism,tasaki1995ferromagnetism,tasaki2003ferromagnetism,tasaki1994stability,tasaki1996stability,tanaka2020extension}.
The highly restricted form of this configuration allows us to represent the ground state using only the information of the kinetic Hamiltonian, considerably simplifying the problem~\cite{tasaki2020physics}.
For its description, let us consider the following repulsive Hubbard model
\ba
H=\sum_{\bk \al \be \sg} h(\bk)_{\al \be} \cdag_{\bk \al \sg} c_{\bk \be \sg} + U\sum_{\bR \al} n_{\bR \al \ua} n_{\bR \al \da},
\label{eq:hubbard}
\ea
where $h(\bk)$ and $U>0$ indicate the kinetic Hamiltonian and the local Hubbard repulsion, respectively.
$c_{\bk \al \sg}$ denotes the annihilation operator of an electron with momentum $\bk$, orbital index $\al = 1, \dots, \norb$, and spin $\sg = \ua, \da$.
$n_{\bR \al \sg} \equiv \cdag_{\bR \al \sg} c_{\bR \al \sg}$ is the corresponding number operator, and $\bR = \bR_1, \dots, \bR_{\nc}$ denotes the $\nc$ unit cells.
In addition, we fix the electron number to $\ntot = \nc \nocc$, and assume that upon filling one spin sector, $h(\bk)$ exhibits a momentum-independent band filling $\nocc$~\cite{supple}.
Let us denote the local and total spin operators as $\bS_{\bR \al} = \hf \sum_{\sigma_1,\sigma_2}\cdag_{\bR \al \sg_1}\boldsymbol{\sg}_{\sg_1 \sg_2} c_{\bR \al \sg_2}$ and
$\bS_{tot} = \sum_{\bR \al} \bS_{\bR \al}$, respectively ($\boldsymbol{\sg}$ is the Pauli matrix, and we set $\hbar=1$).
The system is said to exhibit SFM if and only if
\ba
(\bS_{tot})^2 \ket{GS} = \smax (\smax + 1) \ket{GS}
\quad
(\smax \equiv \frac{\ntot}{2})
\label{seq:sat}
\ea
for any ground state $\ket{GS}$~\cite{tasaki2020physics}.
Denoting the single-particle eigensystem as
\ba
\pdag_{n \bk \sg}=\sum_{\al}^{\norb} \cdag_{\bk \al \sg} \ket{u_n(\bk)}_{\al},
~
h(\bk) \ket{u_n(\bk)} = E_n(\bk) \ket{u_n(\bk)},
\ea
where $\ket{u_n(\bk)}_{\al}$ is the $\al$-th component of the $\norb \times 1$ column vector $\ket{u_n(\bk)}$, the ground state manifold is fully determined by performing SU(2) rotations to the fully polarized state $\gs = \prod_{n}^{\nocc} \prod_{\bk} \pdag_{n \bk \ua} \vac$, where $\vac$ is the vacuum state.
Being a single Slater determinant, mean-field analysis on $\gs$ is exact for any $U$, which allows us to rigorously investigate the strongly correlated regime in the following~\cite{supple}.
\textit{Spin excitations.---}
Let us describe the spin excitation spectrum of saturated ferromagnets, generally composed of the Stoner continuum and the low-energy magnon modes~\cite{kusakabe1994ferromagnetic,solyom2010fundamentals}.
We begin from the saturated ferromagnetic state $\gs$ and denote an arbitrary spin excitation that lowers $S_{tot}^z$ by $1$ and carries momentum $\bQ$ as
\ba
\ket{\bQ}=\sum_{\bk}^{\nc} \sum_{\al}^{\norb} \sum_n^{\nocc} z_{\bk \al n}(\bQ) \cdag_{\bk+\bQ \al \da} \psi_{n \bk \ua} \gs,
\label{eq:magnon}
\ea
where  $z_{\bk \al n}(\bQ)$ is an arbitrary number.
We define the spin excitation energy as
\ba
\Emag = \frac{\braq H \ketq}{\normsq{\bQ}}-\bra{\Phi} H \ket{\Phi},
\label{eq:var}
\ea
and decouple the many-body terms in \eq{eq:var} using mean-field theory and apply the variational principle to minimize $\Emag$ with respect to $\zst$~\cite{shankar2012principles,griffiths2018introduction,wu2020quantum}.
This gives the following eigenvalue equation
\ba
\sum_{\bk' \be n'} \Heff z_{\bk' \be n'}(\bQ)
=
\Emag z_{\bk \al n}(\bQ),
\label{eq:varsol}
\ea
where the spin excitation Hamiltonian $\Hse (\bQ)$ is an $\nc \norb \nocc \times \nc \norb \nocc$ Hermitian matrix given by~\cite{supple}
\ba
\Heff
&=
[h(\bk+\bQ)_{\al \be}
-
E_n(\bk)\delta_{\al \be}] \delta_{\bk \bk'} \delta_{n n'}
\nn
&+
\Uc[ \sum_{\bq}^{\nc} \sum_l^{\nocc} \ls{\al} \brkr{u_l(\bq)}{u_l(\bq)}_{\al} \delta_{\bk \bk'} \delta_{n n'}
\nn
&-
\ls{\al} \brkr{u_{n'}(\bk')}{u_n(\bk)}_{\al}] \delta_{\al \be}.
\label{eq:Heff}
\ea
In the Supplementary Materials (SM), we prove that the mean-field decoupling in \eq{eq:var} is exact at any $U > 0$ using Wick's theorem~\cite{wick1950evaluation,giuliani2008quantum}.

\begin{figure}[t]
\centering\includegraphics[width=0.48\textwidth]{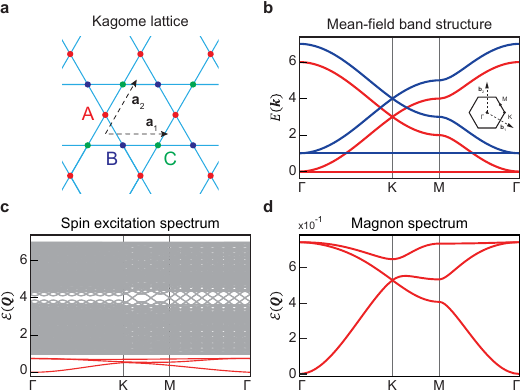}
\caption{{\bf Spin excitation spectrum of the kagom\'{e} ferromagnet.}
	(a) The kagom\'{e} lattice.
	(b) The mean-field band structure for $\nocc = 1$, $t = 1$, $U = 3$, and $\mu=-2$ from \eq{eq:kagome}, where the lowest flat band in red is fully occupied. Red and blue lines represent the $\ua$ and $\da$ spin bands, respectively. 
	(c) The spin excitation spectrum. Red and grey lines correspond to the magnon bands and Stoner continuum, respectively. 
	(d) The magnon band structure.}
\label{fig1}
\end{figure}
%

%
For illustration, let us describe the spin excitation spectrum of a saturated ferromagnet on the kagom\'{e} lattice shown in \fig{fig1} (a).
The tight-binding Hamiltonian containing only the nearest-neighbor hopping amplitude $t>0$ is given by
\ba
h(\bk) = 2t
\bpm
0 & \cos \hfx{\bk \cdot \bba_1}  & \cos \hfx{\bk \cdot \bba_2} \\
\cos \hfx{\bk \cdot \bba_1} & 0 & \cos \hfx{\bk \cdot (\bba_1 - \bba_2)} \\
\cos \hfx{\bk \cdot \bba_2} & \cos \hfx{\bk \cdot (\bba_1 - \bba_2)} & 0
\epm
-\mu \id,
\label{eq:kagome}
\ea
where $\bb a_1 = (1,0)$ and $\bb a_2 = (\hf, \frac{\sqrt{3}}{2})$, $\mu$ is the chemical potential, and $\id$ is an identity matrix.
This model has a spin-degenerate flat band at the bottom, and exhibits SFM at any $U > 0$ when the flat band is half filled with ($\nocc = 1$)~\cite{mielke1991ferromagnetic,mielke1992exact,half}.
The corresponding spin-split mean-field band structure is shown in \fig{fig1} (b).
The relevant spin excitation spectrum in \fig{fig1}(c), composed of the Stoner continuum and spin wave excitations, is obtained by diagonalizing \eq{eq:Heff} at each $\bQ$.
In general, there are $\norb$ magnons, which consists of $1$ gapless Goldstone mode and $\norb - 1$ gapped modes.
When the ground state is ferromagnetic, the  Goldstone mode exhibits a quadratic dispersion with an energy minimum at $\bQ = \bb 0$.
For this reason, a positive energy in the spin excitation spectrum is regarded as a strong indicator for stable ferromagnetism~\cite{kusakabe1994ferromagnetic,tasaki1994stability}.
Conversely, a negative (or zero) magnon energy indicates that the ferromagnetic state $\gs$ is unstable~\cite{alavirad2020ferromagnetism,shastry1990instability}.
%

\textit{Approximation to the Goldstone mode.---}
Assuming that $\gs$ is the unique ground state up to SU(2) rotations, we use first order perturbation theory to obtain an analytic expression of the Goldstone mode.
We reorganize \eq{eq:Heff} as $\Hse(\bQ) = \Hse(\bb 0) + V(\bQ)$
and treat $V(\bQ) \equiv \Hse (\bQ) - \Hse (\bb 0)$ as the perturbed Hamiltonian.
Due to the SU(2) symmetry of $H$, $\Hse(\bb 0)$ is exactly solved at $\bQ = 0$ for the Goldstone mode by $z_{\bk' \beta n'}(\bb 0) = \ket{u_{n'} (\bk')}_{\beta}$ with $\mc{E}_G(\bb 0) = 0$.
The Goldstone mode energy up to first order perturbation is given by $\Epert = \frac{\bra{z^G(\bb 0)} V(\bQ) \ket{z^G(\bb 0)}}{\brk{z^{G}(\bb 0)}{z^{G}(\bb 0)}}$, with
\ba
\Epert
=
\den{\ntot}
\sum_{\bk}
Tr[(h(\bk + \bQ)-h(\bk))P(\bk)],
\label{eq:pert}
\ea
where $P(\bk) = \sum_n \kbr{u_n(\bk)}{u_n(\bk)}$ is the occupied band projector~\cite{supple}.
Note that the ambiguity in choosing $P(\bk)$ may be avoided at band crossings by shifting the momentum space grid, or using the gauge choice that returns maximally localized Wannier functions (MLWF) for entangled bands~\cite{souza2001maximally}.
Importantly, \eq{eq:pert} is fully determined by $h(\bk)$, and is independent of the interaction strength.
To understand this behavior, note that perturbation theory is valid when the unperturbed Hamiltonian $\Hse(\bb 0)$ sufficiently larger than the perturbation $V(\bQ)$.
Since all the $U$-dependent terms are in the former, our perturbative analysis describes $\mc{E}_G(\bQ)$ in the strongly correlated limit, that is, $U\rightarrow \infty$.
Moreover, as first order perturbation uses the ground state of the unperturbed Hamiltonian to calculate the energy of the perturbed Hamiltonian, it places a rigorous upper bound on the true ground state energy at any $U$~\cite{supple}.
Thus, we obtain the following inequality
\ba
\Eg|_{U < \infty} \leq \Eg|_{U \rightarrow \infty} \leq \Epert.
\label{eq:ineq}
\ea

\begin{figure}[t]
\centering\includegraphics[width=0.48\textwidth]{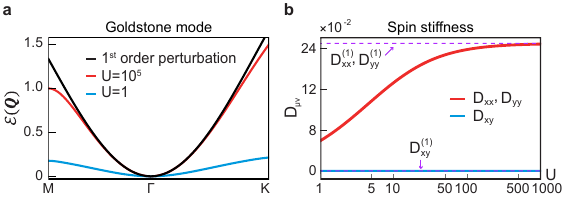}
\caption{{\bf Upper bound of the Goldstone mode of the kagom\'{e} saturated ferromagnet.}
	(a) The Goldstone mode dispersion. Red and blue lines correspond to $U = 10^5$ and $U = 1$, respectively. The black curve indicates the upper bound $\Epert$. 
	(b) Spin stiffness $\dm \dn \Eg|_{\bQ = \bb 0}$ as a function of $U$. Due to $C_{3z}$ symmetry, $D_{xx} = D_{yy}$ and $D_{xy} = 0$. The red and blue curves correspond to $D_{xx} (D_{yy})$ and $D_{xy}$, respectively. The purple dashed lines obtained from $\dm \dn \Epert|_{\bQ = \bb 0}$ are identical to the true spin stiffness at $U \rightarrow \infty$.
}
\label{fig2}
\end{figure}

%

Let us comment on the validity of this approximation.
The many-body spectrum, and consequently, the spin-excitation spectrum, is invariant upon shifting the orbital positions while retaining the hopping amplitudes.
However, such transformation affects $\Epert$~\cite{supple}.
To obtain the optimal approximation, we enforce the minimal metric condition~\cite{huhtinen2022revisiting,herzog2022many}, under which the lower bound of the Wannier function spreading is minimized by varying the orbital positions.
Interestingly, in all the examples we considered, the upper bound $\Epert$ saturates in this position choice, as illustrated in the kagom\'{e} lattice model in \fig{fig2}.
Even if this were not the case, $\Epert > 0$ implies $\Eg|_{U \rightarrow \infty}$ unless the second order perturbation has a larger magnitude than the first order term.
Thus, $\Epert$ accurately diagnoses the stability of ferromagnetism for $U \rightarrow \infty$ given that the perturbative framework does not break down.

\textit{Relation to quantum geometry.---}
Let us discuss the geometric aspects of \eq{eq:pert}.
In general, the geometry of the quantum states, or more generally, their projectors, is characterized by the Abelian quantum geometric tensor (QGT) $\qgt$~\cite{cheng2010quantum}, where $\mu, \nu = x, y, \dots$ indicate orthogonal coordinates.
The real and imaginary parts of the QGT are related to the quantum metric~\cite{provost1980riemannian,resta2011insulating} and Berry curvature~\cite{berry1984quantal} by
\ba
Re[\qgt] = \qm, \quad
Im[\qgt] = -\hf \bcurv.
\ea
In particular, the quantum metric defines the local geometry imposed by the Hilbert-Schmidt quantum distance
\ba
ds^2 = \qm dk_{\mu} dk_{\nu},
\quad
s(\bk, \bk')^2 = \nocc - Tr[P(\bk) P(\bk')],
\label{eq:metric}
\ea
which is $\nocc$ ($0$) for orthogonal (identical) projectors~\cite{supple}, thereby introducing a natural geometric measure for the similarity between projectors.
The geometric interpretation is as follows.
In the SM, we prove that $\Epert \geq 0$.
To see this, note that $Tr[h(\bk) P(\bk)]=\sum_n^{\nocc} E_n(\bk)$.
On the other hand, $Tr[h(\bk + \bQ)P(\bk)]$ represents the sum of the expectation values of $h(\bk + \bQ)$ evaluated by the occupied eigenvectors $\ket{u_n(\bk)}$, instead of $\ket{u_n(\bk + \bQ)}$.
Therefore, one obtains
\ba
\Epert \geq \frac{1}{\ntot} \sum_{n \bk}(E_n(\bk + \bQ)-E_n(\bk)) = 0,
\ea
where the equality saturates if and only if $P(\bk) = P(\bk + \bQ)$ for every $\bk$ at a given $\bQ$.
In terms of quantum geometry, this means that $\Epert > 0$ if and only if
\ba
f(\bQ) \equiv \sum_{\bk} s(\bk, \bk + \bQ)^2 > 0.
\ea
Since $f(\bb 0) = 0$, the stability condition at $\Gamma$ requires that the quantum metric integral $G_{\mu\nu} \equiv \frac{1}{\nc} \sum_{\bk} \qm$ is a positive definite tensor.
In turn, a large quantum metric naturally enhances the stability of the Goldstone mode.
Furthermore, we obtain the spin stiffness $\Dmag \equiv \der_{\mu} \der_{\nu} \Em |_{\bQ = \bb 0}$, which is the inverse mass of the Goldstone boson at $\bQ = \bb 0$, as~\cite{supple}
\ba
\Dpert
&=
\frac{1}{\ntot} \sum_{\bk} [\sum_{n \leq \nocc} \dm \dn E_n (\bk) +
\nn
&
\sum_{m>\nocc}\sum_{n \leq\nocc} (E_m (\bk) - E_n (\bk)) (\fid + \chi_{\nu\mu}^{mn}(\bk))],
\label{eq:dpert}
\ea
where $\fid \equiv \brkr{\dm \um}{\un} \brkr{\un}{\dn \um}$ is the \textit{band-resolved quantum geometric tensor}~\cite{watanabe2021chiral,kitamura2022quantum}, whose sum reduces to the QGT as $\qgt
=\sum_{m>\nocc} \sum_{n \leq \nocc} \chi_{\mu\nu}^{nm}(\bk)$.
Being the product of interband Berry connections, $\fid$ is closely related to quantum geometry~\cite{hwang2021geometric}.
\eq{eq:dpert} explicitly shows that $\Epert$ is composed of two separate terms.
One is the sum of the electronic band curvature and the other is the sum of the band-resolved QGTs between the unoccupied and occupied bands weighted by their energy difference.
Note that the first term vanishes if $E_n (\bk)$ are smooth and periodic in the Brillouin zone, which is always the case for insulators; in such cases, the geometric term fully determines the spin stiffness.
Moreover, even in semimetals where the first term does not vanish, there is always a geometric contribution.
This is because a kink in the band structure always induces a singularity in the wave functions, leading to a diverging $\fid$~\cite{supple,hwang2021wave}.
These results collectively show the prominent role of quantum geometry in stabilizing the Goldstone mode in the strongly correlated limit.
We can intuitively understand the role of quantum geometry by considering Heisenberg's direct exchange interaction between two electrons, which minimizes the Coulomb repulsion for spin triplet states for localized wave functions which have a finite overlap~\cite{Heisenberg1928,tasaki1992ferromagnetism,repellin2020ferromagnetism}.
Substituting the localized wave functions with the MLWFs, one can naturally think that a larger Wannier function spreading will result in a stable ferromagnetic ground state in the strongly correlated limit.
Crucially, $G_{\mu\mu}$ is the lower bound of the spreading of the MLWF in the $\mu$-th direction; conversely, the MLWFs can be perfectly localized in the $\mu$-th direction when $G_{\mu\mu} = 0$~\cite{marzari1997maximally}.
Therefore, our result shows that quantum geometry provides a natural measure of Heisenberg's direct exchange interaction, which is necessary for stabilizing SFM in integer-filled insulators or semimetals.
%

\begin{figure}[t]
	\centering\includegraphics[width=.48\textwidth]{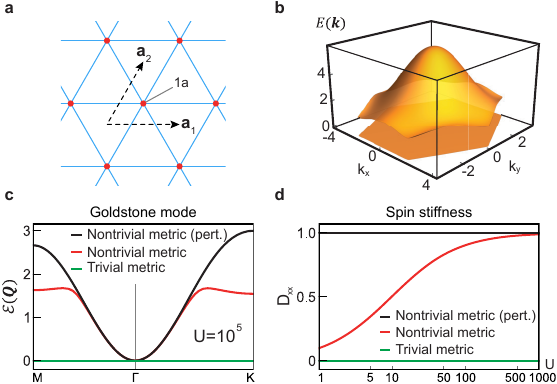}
	\caption{{\bf Influence of the quantum metric on the spin stiffness.}
		(a) Lattice structure belonging to wallpaper group $p3$. $\bba_1 = (1,0)$ and $\bba_2 = (\hf,\hfx{\sqrt{3}})$ are the primitive lattice vectors. 
		(b) The band structure of the Hamiltonian in \eq{eq:oai}. The system consists of a single occupied flat band at $E =0$, which is separated from two degenerate dispersive bands. 
		(c) Goldstone mode dispersion for $U = 10^5$. (d) $D_{xx}$ plotted as a function of $U$. In this model, $D_{xx} = D_{yy}$ and $D_{xy} = 0$ due to $C_{3z}$ symmetry.  We note that $\Dmag (U \rightarrow \infty) = \Dpert$. 
		In (c) and (d), the red and black curves are numerical data and the upper bound, respectively, for the model with nontrivial quantum metric. The green curves are calculated for the case with zero quantum metric. 
	}
	\label{fig3}
\end{figure}

\textit{No-go theorem.---}
We present a no-go theorem that rigorously forbids SFM when $G_{\mu\nu}=0$.
Notably, this occurs for any half-filled Hubbard model, regardless of the form of $h(\bk)$.
This is because the Wannier functions of the occupied bands can always be perfectly localized to the atomic orbitals, resulting in a vanishing direct exchange interaction.
This explains the previous studies on half-filled Hubbard models, where in no circumstance has SFM been observed~\cite{lieb1989two,fazekas1990ground,lee2005u,szasz2020chiral}.
To demonstrate the influence of quantum geometry on the stability of SFM, we calculate the gapless magnon spectra of flat band models with identical energy dispersion, but different quantum metrics.
Explicitly, let us consider a system in the wallpaper group $p3$ as in \fig{fig3}(a), and place three orbitals whose $C_{3z}$ eigenvalues are $1$, $\om$, and $\om^2$ ($\om = \ee{\frac{2 \pi i}{3}}$), respectively, at the $1a$ Wyckoff position.
Here, $C_{3z}$ indicates a 3-fold rotation about the $z$-axis.
We consider one occupied flat band at the Fermi level ($E=0$) and two degenerate unoccupied bands with energy $E_{unocc}(\bk) = 3 + \cos \bk \cdot \bba_1 + \cos \bk \cdot \bba_2 + \cos \bk \cdot (\bba_1 - \bba_2)$ (\fig{fig3}(b)).
The relevant Hamiltonian is given by
\ba
h(\bk) = E_{unocc} (\bk) (\id-P(\bk)),
\label{eq:oai}
\ea
where $P(\bk) = \kbr{u_1 (\bk)}{u_1 (\bk)}$ is the occupied band projector.
In these types of systems where the occupied bands are smooth and the occupied and unoccupied bands are fully degenerate among each other, the spin stiffness reduces to
\ba
\Dpert
&=
\frac{2}{\ntot} \sum_{\bk}^{\nc} (E_m (\bk) - E_n (\bk)) \qm.
\label{eq:twoband}
\ea
Since the quantum metric is given by $\qm = \hf Tr [\der_\mu P(\bk) \der_\nu P(\bk)]$, we can tune $\qm$ by changing $\ket{u_1 (\bk)}$.
The model with zero quantum metric is constructed by using $\ket{u_1 (\bk)} = (1,0,0)^T$.
Following Ref.~\cite{herzog2022superfluid}, we construct the geometrically nontrivial model with
\ba
\ket{u_1 (\bk)} = \frac{1}{3}
\bpm
1 + \ee{-i \bk \cdot \bba_1} + \ee{-i \bk \cdot \bba_2} \\
1 + \om \ee{-i \bk \cdot \bba_1} + \om^2 \ee{-i \bk \cdot \bba_2} \\
1 + \om^2 \ee{-i \bk \cdot \bba_1} + \om \ee{-i \bk \cdot \bba_2}
\epm
,
\ea
whose quantum metric is given by $\qm = \delta_{\mu \nu}/6$.
From \eq{eq:twoband}, we obtain $\Dpert = 0$ and $\delta_{\mu \nu}$ for the trivial and nontrivial models, respectively.
\fig{fig3}(c) and (d) show the Goldstone mode dispersion and the spin stiffness. 
The two models show drastically different behavior despite having the same energy dispersion, which demonstrates the significance of quantum metric on the computed quantities.
Crucially, this shows that even in flat band systems which tend to exhibit SFM~\cite{mielke1991ferromagnetic,mielke1992exact,mielke1993ferromagnetism,mielke1999stability,tasaki1992ferromagnetism,tasaki1995ferromagnetism,tasaki2003ferromagnetism,tasaki1994stability,tasaki1996stability,tanaka2020extension}, the existence of a nontrivial quantum metric is required for such behavior.
%

\textit{Discussion.---}
We have shown that quantum geometry plays an essential role in stabilizing SFM for ferromagnetic insulators or semimetals~\cite{mcguire2015coupling,song2006giant,meng2018strain,zhang2016magnetic,liu2018giant,kim2018large,jin2017ferromagnetic,liu2019magnetic}.
This shows that even in systems completely unrelated to the extensively studied flat band models~\cite{mielke1991ferromagnetic,mielke1992exact,mielke1993ferromagnetism,mielke1999stability,tasaki1992ferromagnetism,tasaki1995ferromagnetism,tasaki2003ferromagnetism,tasaki1994stability,tasaki1996stability,tanaka2020extension,lieb1989two}, quantum geometry may provide a novel route for achieving ferromagnetism, which can potentially serve as a starting point for ferromagnetic materials search from first principles studies.
We emphasize that the necessity of quantum geometry for achieving stable SFM is an exact result, valid for any Hamiltonian with integer occupation, regardless of the spatial dimension or complexity of $h(\bk)$.
Nevertheless, one must consider the competition between different symmetry broken phases to fully characterize the ground state.

An interesting point can be made when one relates the quantum metric to various topological constraints.
For instance, the well-known inequality between the trace of the quantum metric and the Berry curvature~\cite{peotta2015superfluidity,xie2020topology} gives a lower bound of the quantum metric imposed by the Chern number of occupied bands, indicating that the band topology may stabilize SFM, as we demonstrate in the SM.
Similarly, fragile or obstructed atomic band topology also provides a lower bound that can be diagnosed by real space invariants developed recently~\cite{song2020twisted,herzog2022superfluid}.
This shows that the existence of conducting edge states is not the only consequence of topology, in that it may influence the many-body ground state upon introducing Hubbard interaction.
Finally, we note that our theory can be directly applied to pseudospin ferromagnetism~\cite{jung2011lattice,bernevig2021twisted}.
Further extensions to other broken symmetry ground states in itinerant electronic systems would be one important direction for future study~\cite{han2024quantum}.


\begin{acknowledgments}
We thank Seung-Hun Lee and Jonah-Herzog Arbeitman for fruitful discussions regarding this work.
J.K. and B.J.Y. were supported by 
Samsung  Science and Technology Foundation under Project No. SSTF-BA2002-06,
the National Research Foundation of Korea (NRF) grants funded by the government of Korea (MSIT) (Grants No. NRF-2021R1A5A1032996), and GRDC (Global Research Development Center) Cooperative Hub Program through the National Research Foundation of Korea (NRF) funded by the Ministry of Science and ICT (MSIT) (RS-2023-00258359). T.O. was supported by JSPS KAKENHI Grant Numbers 24H00197 and 24H02231.
\end{acknowledgments}

\bibliography{Refs.bib}

\begin{thebibliography}{84}%
\makeatletter
\providecommand \@ifxundefined [1]{%
 \@ifx{#1\undefined}
}%
\providecommand \@ifnum [1]{%
 \ifnum #1\expandafter \@firstoftwo
 \else \expandafter \@secondoftwo
 \fi
}%
\providecommand \@ifx [1]{%
 \ifx #1\expandafter \@firstoftwo
 \else \expandafter \@secondoftwo
 \fi
}%
\providecommand \natexlab [1]{#1}%
\providecommand \enquote  [1]{``#1''}%
\providecommand \bibnamefont  [1]{#1}%
\providecommand \bibfnamefont [1]{#1}%
\providecommand \citenamefont [1]{#1}%
\providecommand \href@noop [0]{\@secondoftwo}%
\providecommand \href [0]{\begingroup \@sanitize@url \@href}%
\providecommand \@href[1]{\@@startlink{#1}\@@href}%
\providecommand \@@href[1]{\endgroup#1\@@endlink}%
\providecommand \@sanitize@url [0]{\catcode `\\12\catcode `\$12\catcode `\&12\catcode `\#12\catcode `\^12\catcode `\_12\catcode `\%12\relax}%
\providecommand \@@startlink[1]{}%
\providecommand \@@endlink[0]{}%
\providecommand \url  [0]{\begingroup\@sanitize@url \@url }%
\providecommand \@url [1]{\endgroup\@href {#1}{\urlprefix }}%
\providecommand \urlprefix  [0]{URL }%
\providecommand \Eprint [0]{\href }%
\providecommand \doibase [0]{http://dx.doi.org/}%
\providecommand \selectlanguage [0]{\@gobble}%
\providecommand \bibinfo  [0]{\@secondoftwo}%
\providecommand \bibfield  [0]{\@secondoftwo}%
\providecommand \translation [1]{[#1]}%
\providecommand \BibitemOpen [0]{}%
\providecommand \bibitemStop [0]{}%
\providecommand \bibitemNoStop [0]{.\EOS\space}%
\providecommand \EOS [0]{\spacefactor3000\relax}%
\providecommand \BibitemShut  [1]{\csname bibitem#1\endcsname}%
\let\auto@bib@innerbib\@empty
\bibitem [{\citenamefont {Heisenberg}(1928)}]{Heisenberg1928}%
  \BibitemOpen
  \bibfield  {author} {\bibinfo {author} {\bibfnamefont {W.}~\bibnamefont {Heisenberg}},\ }\bibfield  {title} {\enquote {\bibinfo {title} {{Zur} {Theorie} des {Ferromagnetismus}},}\ }\href {\doibase 10.1007/bf01328601} {\bibfield  {journal} {\bibinfo  {journal} {Zeitschrift f{\"{u}}r Physik}\ }\textbf {\bibinfo {volume} {49}},\ \bibinfo {pages} {619--636} (\bibinfo {year} {1928})}\BibitemShut {NoStop}%
\bibitem [{\citenamefont {Bloch}(1929)}]{bloch1929bemerkung}%
  \BibitemOpen
  \bibfield  {author} {\bibinfo {author} {\bibfnamefont {Felix}\ \bibnamefont {Bloch}},\ }\bibfield  {title} {\enquote {\bibinfo {title} {{Bemerkung} zur {Elektronentheorie} des {Ferromagnetismus} und der elektrischen {Leitf{\"a}}higkeit},}\ }\href {\doibase 10.1007/BF01340281} {\bibfield  {journal} {\bibinfo  {journal} {Zeitschrift f{\"u}r Physik}\ }\textbf {\bibinfo {volume} {57}},\ \bibinfo {pages} {545--555} (\bibinfo {year} {1929})}\BibitemShut {NoStop}%
\bibitem [{\citenamefont {Gutzwiller}(1963)}]{gutzwiller1963effect}%
  \BibitemOpen
  \bibfield  {author} {\bibinfo {author} {\bibfnamefont {Martin~C}\ \bibnamefont {Gutzwiller}},\ }\bibfield  {title} {\enquote {\bibinfo {title} {{Effect} of correlation on the ferromagnetism of transition metals},}\ }\href {\doibase 10.1103/PhysRevLett.10.159} {\bibfield  {journal} {\bibinfo  {journal} {Physical Review Letters}\ }\textbf {\bibinfo {volume} {10}},\ \bibinfo {pages} {159} (\bibinfo {year} {1963})}\BibitemShut {NoStop}%
\bibitem [{\citenamefont {Kanamori}(1963)}]{kanamori1963electron}%
  \BibitemOpen
  \bibfield  {author} {\bibinfo {author} {\bibfnamefont {Junjiro}\ \bibnamefont {Kanamori}},\ }\bibfield  {title} {\enquote {\bibinfo {title} {{Electron} correlation and ferromagnetism of transition metals},}\ }\href {\doibase 10.1143/PTP.30.275} {\bibfield  {journal} {\bibinfo  {journal} {Progress of Theoretical Physics}\ }\textbf {\bibinfo {volume} {30}},\ \bibinfo {pages} {275--289} (\bibinfo {year} {1963})}\BibitemShut {NoStop}%
\bibitem [{\citenamefont {Hubbard}(1963)}]{hubbard1963electron}%
  \BibitemOpen
  \bibfield  {author} {\bibinfo {author} {\bibfnamefont {John}\ \bibnamefont {Hubbard}},\ }\bibfield  {title} {\enquote {\bibinfo {title} {{Electron} correlations in narrow energy bands},}\ }\href {\doibase 10.1098/rspa.1963.0204} {\bibfield  {journal} {\bibinfo  {journal} {Proceedings of the Royal Society of London. Series A. Mathematical and Physical Sciences}\ }\textbf {\bibinfo {volume} {276}},\ \bibinfo {pages} {238--257} (\bibinfo {year} {1963})}\BibitemShut {NoStop}%
\bibitem [{\citenamefont {Nagaoka}(1966)}]{nagaoka1966ferromagnetism}%
  \BibitemOpen
  \bibfield  {author} {\bibinfo {author} {\bibfnamefont {Yosuke}\ \bibnamefont {Nagaoka}},\ }\bibfield  {title} {\enquote {\bibinfo {title} {{Ferromagnetism} in a narrow, almost half-filled {$s$} band},}\ }\href {\doibase 10.1103/PhysRev.147.392} {\bibfield  {journal} {\bibinfo  {journal} {Physical Review}\ }\textbf {\bibinfo {volume} {147}},\ \bibinfo {pages} {392} (\bibinfo {year} {1966})}\BibitemShut {NoStop}%
\bibitem [{\citenamefont {Tasaki}(1989)}]{tasaki1989extension}%
  \BibitemOpen
  \bibfield  {author} {\bibinfo {author} {\bibfnamefont {Hal}\ \bibnamefont {Tasaki}},\ }\bibfield  {title} {\enquote {\bibinfo {title} {{Extension} of {Nagaoka’s} theorem on the large-{U} {Hubbard} model},}\ }\href {\doibase 10.1103/PhysRevB.40.9192} {\bibfield  {journal} {\bibinfo  {journal} {Physical Review B}\ }\textbf {\bibinfo {volume} {40}},\ \bibinfo {pages} {9192} (\bibinfo {year} {1989})}\BibitemShut {NoStop}%
\bibitem [{\citenamefont {Tasaki}(1998)}]{tasaki1998nagaoka}%
  \BibitemOpen
  \bibfield  {author} {\bibinfo {author} {\bibfnamefont {Hal}\ \bibnamefont {Tasaki}},\ }\bibfield  {title} {\enquote {\bibinfo {title} {{From} {Nagaoka's} ferromagnetism to flat-band ferromagnetism and beyond: an introduction to ferromagnetism in the {Hubbard} model},}\ }\href {\doibase 10.1143/PTP.99.489} {\bibfield  {journal} {\bibinfo  {journal} {Progress of theoretical physics}\ }\textbf {\bibinfo {volume} {99}},\ \bibinfo {pages} {489--548} (\bibinfo {year} {1998})}\BibitemShut {NoStop}%
\bibitem [{\citenamefont {Shastry}\ \emph {et~al.}(1990)\citenamefont {Shastry}, \citenamefont {Krishnamurthy},\ and\ \citenamefont {Anderson}}]{shastry1990instability}%
  \BibitemOpen
  \bibfield  {author} {\bibinfo {author} {\bibfnamefont {B.~S.}\ \bibnamefont {Shastry}}, \bibinfo {author} {\bibfnamefont {H.~R.}\ \bibnamefont {Krishnamurthy}}, \ and\ \bibinfo {author} {\bibfnamefont {P.~W.}\ \bibnamefont {Anderson}},\ }\bibfield  {title} {\enquote {\bibinfo {title} {{Instability} of the {Nagaoka} ferromagnetic state of the {U=\ensuremath{\infty}} {Hubbard} model},}\ }\href {\doibase 10.1103/PhysRevB.41.2375} {\bibfield  {journal} {\bibinfo  {journal} {Phys. Rev. B}\ }\textbf {\bibinfo {volume} {41}},\ \bibinfo {pages} {2375--2379} (\bibinfo {year} {1990})}\BibitemShut {NoStop}%
\bibitem [{\citenamefont {Mielke}(1991)}]{mielke1991ferromagnetic}%
  \BibitemOpen
  \bibfield  {author} {\bibinfo {author} {\bibfnamefont {Andreas}\ \bibnamefont {Mielke}},\ }\bibfield  {title} {\enquote {\bibinfo {title} {{Ferromagnetic} ground states for the {Hubbard} model on line graphs},}\ }\href {\doibase 10.1088/0305-4470/24/2/005} {\bibfield  {journal} {\bibinfo  {journal} {Journal of Physics A: Mathematical and General}\ }\textbf {\bibinfo {volume} {24}},\ \bibinfo {pages} {L73} (\bibinfo {year} {1991})}\BibitemShut {NoStop}%
\bibitem [{\citenamefont {Mielke}(1992)}]{mielke1992exact}%
  \BibitemOpen
  \bibfield  {author} {\bibinfo {author} {\bibfnamefont {A}~\bibnamefont {Mielke}},\ }\bibfield  {title} {\enquote {\bibinfo {title} {{Exact} ground states for the {Hubbard} model on the {Kagome} lattice},}\ }\href {\doibase 10.1088/0305-4470/25/16/011} {\bibfield  {journal} {\bibinfo  {journal} {Journal of Physics A: Mathematical and General}\ }\textbf {\bibinfo {volume} {25}},\ \bibinfo {pages} {4335} (\bibinfo {year} {1992})}\BibitemShut {NoStop}%
\bibitem [{\citenamefont {Mielke}\ and\ \citenamefont {Tasaki}(1993)}]{mielke1993ferromagnetism}%
  \BibitemOpen
  \bibfield  {author} {\bibinfo {author} {\bibfnamefont {Andreas}\ \bibnamefont {Mielke}}\ and\ \bibinfo {author} {\bibfnamefont {Hal}\ \bibnamefont {Tasaki}},\ }\bibfield  {title} {\enquote {\bibinfo {title} {{Ferromagnetism} in the {Hubbard} model: {Examples} from models with degenerate single-electron ground states},}\ }\href {\doibase 10.1007/BF02108079} {\bibfield  {journal} {\bibinfo  {journal} {Communications in mathematical physics}\ }\textbf {\bibinfo {volume} {158}},\ \bibinfo {pages} {341--371} (\bibinfo {year} {1993})}\BibitemShut {NoStop}%
\bibitem [{\citenamefont {Mielke}(1999)}]{mielke1999stability}%
  \BibitemOpen
  \bibfield  {author} {\bibinfo {author} {\bibfnamefont {Andreas}\ \bibnamefont {Mielke}},\ }\bibfield  {title} {\enquote {\bibinfo {title} {{Stability} of ferromagnetism in {Hubbard} models with degenerate single-particle ground states},}\ }\href {\doibase 10.1088/0305-4470/32/48/304} {\bibfield  {journal} {\bibinfo  {journal} {Journal of Physics A: Mathematical and General}\ }\textbf {\bibinfo {volume} {32}},\ \bibinfo {pages} {8411} (\bibinfo {year} {1999})}\BibitemShut {NoStop}%
\bibitem [{\citenamefont {Tasaki}(1992)}]{tasaki1992ferromagnetism}%
  \BibitemOpen
  \bibfield  {author} {\bibinfo {author} {\bibfnamefont {Hal}\ \bibnamefont {Tasaki}},\ }\bibfield  {title} {\enquote {\bibinfo {title} {{Ferromagnetism} in the {Hubbard} models with degenerate single-electron ground states},}\ }\href {\doibase 10.1103/PhysRevLett.69.1608} {\bibfield  {journal} {\bibinfo  {journal} {Physical review letters}\ }\textbf {\bibinfo {volume} {69}},\ \bibinfo {pages} {1608} (\bibinfo {year} {1992})}\BibitemShut {NoStop}%
\bibitem [{\citenamefont {Tasaki}(1995)}]{tasaki1995ferromagnetism}%
  \BibitemOpen
  \bibfield  {author} {\bibinfo {author} {\bibfnamefont {Hal}\ \bibnamefont {Tasaki}},\ }\bibfield  {title} {\enquote {\bibinfo {title} {{Ferromagnetism} in {Hubbard} models},}\ }\href {\doibase 10.1103/PhysRevLett.75.4678} {\bibfield  {journal} {\bibinfo  {journal} {Physical review letters}\ }\textbf {\bibinfo {volume} {75}},\ \bibinfo {pages} {4678} (\bibinfo {year} {1995})}\BibitemShut {NoStop}%
\bibitem [{\citenamefont {Tasaki}(2003)}]{tasaki2003ferromagnetism}%
  \BibitemOpen
  \bibfield  {author} {\bibinfo {author} {\bibfnamefont {Hal}\ \bibnamefont {Tasaki}},\ }\bibfield  {title} {\enquote {\bibinfo {title} {{Ferromagnetism} in the {Hubbard} model: a constructive approach},}\ }\href {\doibase 10.1007/s00220-003-0952-z} {\bibfield  {journal} {\bibinfo  {journal} {Communications in mathematical physics}\ }\textbf {\bibinfo {volume} {242}},\ \bibinfo {pages} {445--472} (\bibinfo {year} {2003})}\BibitemShut {NoStop}%
\bibitem [{\citenamefont {Tasaki}(1994)}]{tasaki1994stability}%
  \BibitemOpen
  \bibfield  {author} {\bibinfo {author} {\bibfnamefont {Hal}\ \bibnamefont {Tasaki}},\ }\bibfield  {title} {\enquote {\bibinfo {title} {{Stability} of ferromagnetism in the {Hubbard} model},}\ }\href {\doibase 10.1103/PhysRevLett.73.1158} {\bibfield  {journal} {\bibinfo  {journal} {Physical review letters}\ }\textbf {\bibinfo {volume} {73}},\ \bibinfo {pages} {1158} (\bibinfo {year} {1994})}\BibitemShut {NoStop}%
\bibitem [{\citenamefont {Tasaki}(1996)}]{tasaki1996stability}%
  \BibitemOpen
  \bibfield  {author} {\bibinfo {author} {\bibfnamefont {Hal}\ \bibnamefont {Tasaki}},\ }\bibfield  {title} {\enquote {\bibinfo {title} {{Stability} of ferromagnetism in {Hubbard} models with nearly flat bands},}\ }\href {\doibase 10.1007/BF02179652} {\bibfield  {journal} {\bibinfo  {journal} {Journal of statistical physics}\ }\textbf {\bibinfo {volume} {84}},\ \bibinfo {pages} {535--653} (\bibinfo {year} {1996})}\BibitemShut {NoStop}%
\bibitem [{\citenamefont {Tanaka}(2020)}]{tanaka2020extension}%
  \BibitemOpen
  \bibfield  {author} {\bibinfo {author} {\bibfnamefont {Akinori}\ \bibnamefont {Tanaka}},\ }\bibfield  {title} {\enquote {\bibinfo {title} {{An} extension of the cell-construction method for the flat-band ferromagnetism},}\ }\href {\doibase 10.1007/s10955-020-02610-3} {\bibfield  {journal} {\bibinfo  {journal} {Journal of Statistical Physics}\ }\textbf {\bibinfo {volume} {181}},\ \bibinfo {pages} {897--916} (\bibinfo {year} {2020})}\BibitemShut {NoStop}%
\bibitem [{\citenamefont {Rhim}\ \emph {et~al.}(2020)\citenamefont {Rhim}, \citenamefont {Kim},\ and\ \citenamefont {Yang}}]{rhim2020quantum}%
  \BibitemOpen
  \bibfield  {author} {\bibinfo {author} {\bibfnamefont {Jun-Won}\ \bibnamefont {Rhim}}, \bibinfo {author} {\bibfnamefont {Kyoo}\ \bibnamefont {Kim}}, \ and\ \bibinfo {author} {\bibfnamefont {Bohm-Jung}\ \bibnamefont {Yang}},\ }\bibfield  {title} {\enquote {\bibinfo {title} {{Quantum} distance and anomalous {Landau} levels of flat bands},}\ }\href {\doibase 10.1038/s41586-020-2540-1} {\bibfield  {journal} {\bibinfo  {journal} {Nature}\ }\textbf {\bibinfo {volume} {584}},\ \bibinfo {pages} {59--63} (\bibinfo {year} {2020})}\BibitemShut {NoStop}%
\bibitem [{\citenamefont {Hwang}\ \emph {et~al.}(2021{\natexlab{a}})\citenamefont {Hwang}, \citenamefont {Rhim},\ and\ \citenamefont {Yang}}]{hwang2021geometric}%
  \BibitemOpen
  \bibfield  {author} {\bibinfo {author} {\bibfnamefont {Yoonseok}\ \bibnamefont {Hwang}}, \bibinfo {author} {\bibfnamefont {Jun-Won}\ \bibnamefont {Rhim}}, \ and\ \bibinfo {author} {\bibfnamefont {Bohm-Jung}\ \bibnamefont {Yang}},\ }\bibfield  {title} {\enquote {\bibinfo {title} {{Geometric} characterization of anomalous {Landau} levels of isolated flat bands},}\ }\href {\doibase 10.1038/s41467-021-26765-z} {\bibfield  {journal} {\bibinfo  {journal} {Nature communications}\ }\textbf {\bibinfo {volume} {12}},\ \bibinfo {pages} {6433} (\bibinfo {year} {2021}{\natexlab{a}})}\BibitemShut {NoStop}%
\bibitem [{\citenamefont {Jung}\ \emph {et~al.}(2024)\citenamefont {Jung}, \citenamefont {Lim},\ and\ \citenamefont {Yang}}]{jung2024quantum}%
  \BibitemOpen
  \bibfield  {author} {\bibinfo {author} {\bibfnamefont {Junseo}\ \bibnamefont {Jung}}, \bibinfo {author} {\bibfnamefont {Hyeongmuk}\ \bibnamefont {Lim}}, \ and\ \bibinfo {author} {\bibfnamefont {Bohm-Jung}\ \bibnamefont {Yang}},\ }\bibfield  {title} {\enquote {\bibinfo {title} {{Quantum} geometry and {Landau} levels of quadratic band crossings},}\ }\href {\doibase 10.1103/PhysRevB.109.035134} {\bibfield  {journal} {\bibinfo  {journal} {Physical Review B}\ }\textbf {\bibinfo {volume} {109}},\ \bibinfo {pages} {035134} (\bibinfo {year} {2024})}\BibitemShut {NoStop}%
\bibitem [{\citenamefont {Liang}\ \emph {et~al.}(2017)\citenamefont {Liang}, \citenamefont {Vanhala}, \citenamefont {Peotta}, \citenamefont {Siro}, \citenamefont {Harju},\ and\ \citenamefont {T{\"o}rm{\"a}}}]{liang2017band}%
  \BibitemOpen
  \bibfield  {author} {\bibinfo {author} {\bibfnamefont {Long}\ \bibnamefont {Liang}}, \bibinfo {author} {\bibfnamefont {Tuomas~I}\ \bibnamefont {Vanhala}}, \bibinfo {author} {\bibfnamefont {Sebastiano}\ \bibnamefont {Peotta}}, \bibinfo {author} {\bibfnamefont {Topi}\ \bibnamefont {Siro}}, \bibinfo {author} {\bibfnamefont {Ari}\ \bibnamefont {Harju}}, \ and\ \bibinfo {author} {\bibfnamefont {P{\"a}ivi}\ \bibnamefont {T{\"o}rm{\"a}}},\ }\bibfield  {title} {\enquote {\bibinfo {title} {{Band} geometry, {Berry} curvature, and superfluid weight},}\ }\href {\doibase 10.1103/PhysRevB.95.024515} {\bibfield  {journal} {\bibinfo  {journal} {Physical Review B}\ }\textbf {\bibinfo {volume} {95}},\ \bibinfo {pages} {024515} (\bibinfo {year} {2017})}\BibitemShut {NoStop}%
\bibitem [{\citenamefont {Xie}\ \emph {et~al.}(2020)\citenamefont {Xie}, \citenamefont {Song}, \citenamefont {Lian},\ and\ \citenamefont {Bernevig}}]{xie2020topology}%
  \BibitemOpen
  \bibfield  {author} {\bibinfo {author} {\bibfnamefont {Fang}\ \bibnamefont {Xie}}, \bibinfo {author} {\bibfnamefont {Zhida}\ \bibnamefont {Song}}, \bibinfo {author} {\bibfnamefont {Biao}\ \bibnamefont {Lian}}, \ and\ \bibinfo {author} {\bibfnamefont {B~Andrei}\ \bibnamefont {Bernevig}},\ }\bibfield  {title} {\enquote {\bibinfo {title} {{Topology-bounded} superfluid weight in twisted bilayer graphene},}\ }\href {\doibase 10.1103/PhysRevLett.124.167002} {\bibfield  {journal} {\bibinfo  {journal} {Physical review letters}\ }\textbf {\bibinfo {volume} {124}},\ \bibinfo {pages} {167002} (\bibinfo {year} {2020})}\BibitemShut {NoStop}%
\bibitem [{\citenamefont {Tian}\ \emph {et~al.}(2023)\citenamefont {Tian}, \citenamefont {Gao}, \citenamefont {Zhang}, \citenamefont {Che}, \citenamefont {Xu}, \citenamefont {Cheung}, \citenamefont {Watanabe}, \citenamefont {Taniguchi}, \citenamefont {Randeria}, \citenamefont {Zhang} \emph {et~al.}}]{tian2023evidence}%
  \BibitemOpen
  \bibfield  {author} {\bibinfo {author} {\bibfnamefont {Haidong}\ \bibnamefont {Tian}}, \bibinfo {author} {\bibfnamefont {Xueshi}\ \bibnamefont {Gao}}, \bibinfo {author} {\bibfnamefont {Yuxin}\ \bibnamefont {Zhang}}, \bibinfo {author} {\bibfnamefont {Shi}\ \bibnamefont {Che}}, \bibinfo {author} {\bibfnamefont {Tianyi}\ \bibnamefont {Xu}}, \bibinfo {author} {\bibfnamefont {Patrick}\ \bibnamefont {Cheung}}, \bibinfo {author} {\bibfnamefont {Kenji}\ \bibnamefont {Watanabe}}, \bibinfo {author} {\bibfnamefont {Takashi}\ \bibnamefont {Taniguchi}}, \bibinfo {author} {\bibfnamefont {Mohit}\ \bibnamefont {Randeria}}, \bibinfo {author} {\bibfnamefont {Fan}\ \bibnamefont {Zhang}},  \emph {et~al.},\ }\bibfield  {title} {\enquote {\bibinfo {title} {{Evidence} for {Dirac} flat band superconductivity enabled by quantum geometry},}\ }\href {\doibase 10.1038/s41586-022-05576-2} {\bibfield  {journal} {\bibinfo  {journal} {Nature}\ }\textbf {\bibinfo {volume} {614}},\ \bibinfo {pages} {440--444} (\bibinfo {year}
  {2023})}\BibitemShut {NoStop}%
\bibitem [{\citenamefont {Gao}\ \emph {et~al.}(2023)\citenamefont {Gao}, \citenamefont {Liu}, \citenamefont {Qiu}, \citenamefont {Ghosh}, \citenamefont {V.~Trevisan}, \citenamefont {Onishi}, \citenamefont {Hu}, \citenamefont {Qian}, \citenamefont {Tien}, \citenamefont {Chen} \emph {et~al.}}]{gao2023quantum}%
  \BibitemOpen
  \bibfield  {author} {\bibinfo {author} {\bibfnamefont {Anyuan}\ \bibnamefont {Gao}}, \bibinfo {author} {\bibfnamefont {Yu-Fei}\ \bibnamefont {Liu}}, \bibinfo {author} {\bibfnamefont {Jian-Xiang}\ \bibnamefont {Qiu}}, \bibinfo {author} {\bibfnamefont {Barun}\ \bibnamefont {Ghosh}}, \bibinfo {author} {\bibfnamefont {Tha{\'\i}s}\ \bibnamefont {V.~Trevisan}}, \bibinfo {author} {\bibfnamefont {Yugo}\ \bibnamefont {Onishi}}, \bibinfo {author} {\bibfnamefont {Chaowei}\ \bibnamefont {Hu}}, \bibinfo {author} {\bibfnamefont {Tiema}\ \bibnamefont {Qian}}, \bibinfo {author} {\bibfnamefont {Hung-Ju}\ \bibnamefont {Tien}}, \bibinfo {author} {\bibfnamefont {Shao-Wen}\ \bibnamefont {Chen}},  \emph {et~al.},\ }\bibfield  {title} {\enquote {\bibinfo {title} {{Quantum} metric nonlinear {Hall} effect in a topological antiferromagnetic heterostructure},}\ }\href {\doibase 10.1126/science.adf1506} {\bibfield  {journal} {\bibinfo  {journal} {Science}\ }\textbf {\bibinfo {volume} {381}},\ \bibinfo {pages} {181--186} (\bibinfo
  {year} {2023})}\BibitemShut {NoStop}%
\bibitem [{\citenamefont {Wang}\ \emph {et~al.}(2023)\citenamefont {Wang}, \citenamefont {Kaplan}, \citenamefont {Zhang}, \citenamefont {Holder}, \citenamefont {Cao}, \citenamefont {Wang}, \citenamefont {Zhou}, \citenamefont {Zhou}, \citenamefont {Jiang}, \citenamefont {Zhang} \emph {et~al.}}]{wang2023quantum}%
  \BibitemOpen
  \bibfield  {author} {\bibinfo {author} {\bibfnamefont {Naizhou}\ \bibnamefont {Wang}}, \bibinfo {author} {\bibfnamefont {Daniel}\ \bibnamefont {Kaplan}}, \bibinfo {author} {\bibfnamefont {Zhaowei}\ \bibnamefont {Zhang}}, \bibinfo {author} {\bibfnamefont {Tobias}\ \bibnamefont {Holder}}, \bibinfo {author} {\bibfnamefont {Ning}\ \bibnamefont {Cao}}, \bibinfo {author} {\bibfnamefont {Aifeng}\ \bibnamefont {Wang}}, \bibinfo {author} {\bibfnamefont {Xiaoyuan}\ \bibnamefont {Zhou}}, \bibinfo {author} {\bibfnamefont {Feifei}\ \bibnamefont {Zhou}}, \bibinfo {author} {\bibfnamefont {Zhengzhi}\ \bibnamefont {Jiang}}, \bibinfo {author} {\bibfnamefont {Chusheng}\ \bibnamefont {Zhang}},  \emph {et~al.},\ }\bibfield  {title} {\enquote {\bibinfo {title} {{Quantum-metric-induced} nonlinear transport in a topological antiferromagnet},}\ }\href {\doibase 10.1038/s41586-023-06363-3} {\bibfield  {journal} {\bibinfo  {journal} {Nature}\ }\textbf {\bibinfo {volume} {621}},\ \bibinfo {pages} {487--492} (\bibinfo {year}
  {2023})}\BibitemShut {NoStop}%
\bibitem [{\citenamefont {Herzog-Arbeitman}\ \emph {et~al.}(2022{\natexlab{a}})\citenamefont {Herzog-Arbeitman}, \citenamefont {Chew}, \citenamefont {Huhtinen}, \citenamefont {T{\"o}rm{\"a}},\ and\ \citenamefont {Bernevig}}]{herzog2022many}%
  \BibitemOpen
  \bibfield  {author} {\bibinfo {author} {\bibfnamefont {Jonah}\ \bibnamefont {Herzog-Arbeitman}}, \bibinfo {author} {\bibfnamefont {Aaron}\ \bibnamefont {Chew}}, \bibinfo {author} {\bibfnamefont {Kukka-Emilia}\ \bibnamefont {Huhtinen}}, \bibinfo {author} {\bibfnamefont {P{\"a}ivi}\ \bibnamefont {T{\"o}rm{\"a}}}, \ and\ \bibinfo {author} {\bibfnamefont {B~Andrei}\ \bibnamefont {Bernevig}},\ }\bibfield  {title} {\enquote {\bibinfo {title} {{Many-body} superconductivity in topological flat bands},}\ }\href {\doibase 10.48550/arXiv.2209.00007} {\bibfield  {journal} {\bibinfo  {journal} {arXiv preprint arXiv:2209.00007}\ } (\bibinfo {year} {2022}{\natexlab{a}}),\ 10.48550/arXiv.2209.00007}\BibitemShut {NoStop}%
\bibitem [{\citenamefont {Han}\ \emph {et~al.}(2024)\citenamefont {Han}, \citenamefont {Herzog-Arbeitman}, \citenamefont {Bernevig},\ and\ \citenamefont {Kivelson}}]{han2024quantum}%
  \BibitemOpen
  \bibfield  {author} {\bibinfo {author} {\bibfnamefont {Zhaoyu}\ \bibnamefont {Han}}, \bibinfo {author} {\bibfnamefont {Jonah}\ \bibnamefont {Herzog-Arbeitman}}, \bibinfo {author} {\bibfnamefont {B~Andrei}\ \bibnamefont {Bernevig}}, \ and\ \bibinfo {author} {\bibfnamefont {Steven~A}\ \bibnamefont {Kivelson}},\ }\bibfield  {title} {\enquote {\bibinfo {title} {{“Quantum} {Geometric} {Nesting”} and {Solvable} {Model} {Flat}-{Band} {Systems}},}\ }\href {\doibase 10.1103/PhysRevX.14.041004} {\bibfield  {journal} {\bibinfo  {journal} {Physical Review X}\ }\textbf {\bibinfo {volume} {14}},\ \bibinfo {pages} {041004} (\bibinfo {year} {2024})}\BibitemShut {NoStop}%
\bibitem [{\citenamefont {Peotta}\ and\ \citenamefont {T{\"o}rm{\"a}}(2015)}]{peotta2015superfluidity}%
  \BibitemOpen
  \bibfield  {author} {\bibinfo {author} {\bibfnamefont {Sebastiano}\ \bibnamefont {Peotta}}\ and\ \bibinfo {author} {\bibfnamefont {P{\"a}ivi}\ \bibnamefont {T{\"o}rm{\"a}}},\ }\bibfield  {title} {\enquote {\bibinfo {title} {{Superfluidity} in topologically nontrivial flat bands},}\ }\href {\doibase 10.1038/ncomms9944} {\bibfield  {journal} {\bibinfo  {journal} {Nature communications}\ }\textbf {\bibinfo {volume} {6}},\ \bibinfo {pages} {8944} (\bibinfo {year} {2015})}\BibitemShut {NoStop}%
\bibitem [{\citenamefont {Hu}\ \emph {et~al.}(2023)\citenamefont {Hu}, \citenamefont {Chen},\ and\ \citenamefont {Law}}]{hu2023anomalous}%
  \BibitemOpen
  \bibfield  {author} {\bibinfo {author} {\bibfnamefont {Jin-Xin}\ \bibnamefont {Hu}}, \bibinfo {author} {\bibfnamefont {Shuai~A}\ \bibnamefont {Chen}}, \ and\ \bibinfo {author} {\bibfnamefont {Kam~Tuen}\ \bibnamefont {Law}},\ }\bibfield  {title} {\enquote {\bibinfo {title} {{Anomalous} coherence length in superconductors with quantum metric},}\ }\href {\doibase 10.48550/arXiv.2308.05686} {\bibfield  {journal} {\bibinfo  {journal} {arXiv preprint arXiv:2308.05686}\ } (\bibinfo {year} {2023}),\ 10.48550/arXiv.2308.05686}\BibitemShut {NoStop}%
\bibitem [{\citenamefont {Chen}\ and\ \citenamefont {Law}(2024)}]{chen2024ginzburg}%
  \BibitemOpen
  \bibfield  {author} {\bibinfo {author} {\bibfnamefont {Shuai~A}\ \bibnamefont {Chen}}\ and\ \bibinfo {author} {\bibfnamefont {KT}~\bibnamefont {Law}},\ }\bibfield  {title} {\enquote {\bibinfo {title} {{Ginzburg-Landau} {Theory} of {Flat-Band} {Superconductors} with {Quantum} {Metric}},}\ }\href {\doibase 10.1103/PhysRevLett.132.026002} {\bibfield  {journal} {\bibinfo  {journal} {Physical Review Letters}\ }\textbf {\bibinfo {volume} {132}},\ \bibinfo {pages} {026002} (\bibinfo {year} {2024})}\BibitemShut {NoStop}%
\bibitem [{\citenamefont {Villegas}\ and\ \citenamefont {Yang}(2021)}]{villegas2021anomalous}%
  \BibitemOpen
  \bibfield  {author} {\bibinfo {author} {\bibfnamefont {Kristian Hauser~A}\ \bibnamefont {Villegas}}\ and\ \bibinfo {author} {\bibfnamefont {Bo}~\bibnamefont {Yang}},\ }\bibfield  {title} {\enquote {\bibinfo {title} {{Anomalous} {Higgs} oscillations mediated by {Berry} curvature and quantum metric},}\ }\href {\doibase 10.1103/PhysRevB.104.L180502} {\bibfield  {journal} {\bibinfo  {journal} {Physical Review B}\ }\textbf {\bibinfo {volume} {104}},\ \bibinfo {pages} {L180502} (\bibinfo {year} {2021})}\BibitemShut {NoStop}%
\bibitem [{\citenamefont {Srivastava}\ and\ \citenamefont {Imamo{\u{g}}lu}(2015)}]{srivastava2015signatures}%
  \BibitemOpen
  \bibfield  {author} {\bibinfo {author} {\bibfnamefont {Ajit}\ \bibnamefont {Srivastava}}\ and\ \bibinfo {author} {\bibfnamefont {Ata{\c{c}}}\ \bibnamefont {Imamo{\u{g}}lu}},\ }\bibfield  {title} {\enquote {\bibinfo {title} {{Signatures} of {Bloch-band} geometry on excitons: nonhydrogenic spectra in transition-metal dichalcogenides},}\ }\href {\doibase 10.1103/PhysRevLett.115.166802} {\bibfield  {journal} {\bibinfo  {journal} {Physical review letters}\ }\textbf {\bibinfo {volume} {115}},\ \bibinfo {pages} {166802} (\bibinfo {year} {2015})}\BibitemShut {NoStop}%
\bibitem [{\citenamefont {Cao}\ \emph {et~al.}(2021)\citenamefont {Cao}, \citenamefont {Fertig},\ and\ \citenamefont {Brey}}]{cao2021quantum}%
  \BibitemOpen
  \bibfield  {author} {\bibinfo {author} {\bibfnamefont {Jinlyu}\ \bibnamefont {Cao}}, \bibinfo {author} {\bibfnamefont {HA}~\bibnamefont {Fertig}}, \ and\ \bibinfo {author} {\bibfnamefont {Luis}\ \bibnamefont {Brey}},\ }\bibfield  {title} {\enquote {\bibinfo {title} {{Quantum} geometric exciton drift velocity},}\ }\href {\doibase 10.1103/PhysRevB.103.115422} {\bibfield  {journal} {\bibinfo  {journal} {Physical Review B}\ }\textbf {\bibinfo {volume} {103}},\ \bibinfo {pages} {115422} (\bibinfo {year} {2021})}\BibitemShut {NoStop}%
\bibitem [{\citenamefont {Hu}\ \emph {et~al.}(2022)\citenamefont {Hu}, \citenamefont {Hyart}, \citenamefont {Pikulin},\ and\ \citenamefont {Rossi}}]{hu2022quantum}%
  \BibitemOpen
  \bibfield  {author} {\bibinfo {author} {\bibfnamefont {Xiang}\ \bibnamefont {Hu}}, \bibinfo {author} {\bibfnamefont {Timo}\ \bibnamefont {Hyart}}, \bibinfo {author} {\bibfnamefont {Dmitry~I}\ \bibnamefont {Pikulin}}, \ and\ \bibinfo {author} {\bibfnamefont {Enrico}\ \bibnamefont {Rossi}},\ }\bibfield  {title} {\enquote {\bibinfo {title} {{Quantum-metric-enabled} exciton condensate in double twisted bilayer graphene},}\ }\href {\doibase 10.1103/PhysRevB.105.L140506} {\bibfield  {journal} {\bibinfo  {journal} {Physical Review B}\ }\textbf {\bibinfo {volume} {105}},\ \bibinfo {pages} {L140506} (\bibinfo {year} {2022})}\BibitemShut {NoStop}%
\bibitem [{\citenamefont {Bernevig}\ \emph {et~al.}(2021)\citenamefont {Bernevig}, \citenamefont {Lian}, \citenamefont {Cowsik}, \citenamefont {Xie}, \citenamefont {Regnault},\ and\ \citenamefont {Song}}]{bernevig2021twisted}%
  \BibitemOpen
  \bibfield  {author} {\bibinfo {author} {\bibfnamefont {B~Andrei}\ \bibnamefont {Bernevig}}, \bibinfo {author} {\bibfnamefont {Biao}\ \bibnamefont {Lian}}, \bibinfo {author} {\bibfnamefont {Aditya}\ \bibnamefont {Cowsik}}, \bibinfo {author} {\bibfnamefont {Fang}\ \bibnamefont {Xie}}, \bibinfo {author} {\bibfnamefont {Nicolas}\ \bibnamefont {Regnault}}, \ and\ \bibinfo {author} {\bibfnamefont {Zhi-Da}\ \bibnamefont {Song}},\ }\bibfield  {title} {\enquote {\bibinfo {title} {{Twisted} bilayer graphene. {V.} {Exact} analytic many-body excitations in {Coulomb} {Hamiltonians:} {Charge} gap, {Goldstone} modes, and absence of {Cooper} pairing},}\ }\href {\doibase 10.1103/PhysRevB.103.205415} {\bibfield  {journal} {\bibinfo  {journal} {Physical Review B}\ }\textbf {\bibinfo {volume} {103}},\ \bibinfo {pages} {205415} (\bibinfo {year} {2021})}\BibitemShut {NoStop}%
\bibitem [{\citenamefont {Wu}\ and\ \citenamefont {Das~Sarma}(2020)}]{wu2020quantum}%
  \BibitemOpen
  \bibfield  {author} {\bibinfo {author} {\bibfnamefont {Fengcheng}\ \bibnamefont {Wu}}\ and\ \bibinfo {author} {\bibfnamefont {S}~\bibnamefont {Das~Sarma}},\ }\bibfield  {title} {\enquote {\bibinfo {title} {{Quantum} geometry and stability of moir{\'e} flatband ferromagnetism},}\ }\href {\doibase 10.1103/PhysRevB.102.165118} {\bibfield  {journal} {\bibinfo  {journal} {Physical Review B}\ }\textbf {\bibinfo {volume} {102}},\ \bibinfo {pages} {165118} (\bibinfo {year} {2020})}\BibitemShut {NoStop}%
\bibitem [{\citenamefont {T{\"o}rm{\"a}}(2023)}]{torma2023essay}%
  \BibitemOpen
  \bibfield  {author} {\bibinfo {author} {\bibfnamefont {P{\"a}ivi}\ \bibnamefont {T{\"o}rm{\"a}}},\ }\bibfield  {title} {\enquote {\bibinfo {title} {{Essay:} {Where} {Can} {Quantum} {Geometry} {Lead} {Us?}}}\ }\href {\doibase 10.1103/PhysRevLett.131.240001} {\bibfield  {journal} {\bibinfo  {journal} {Physical Review Letters}\ }\textbf {\bibinfo {volume} {131}},\ \bibinfo {pages} {240001} (\bibinfo {year} {2023})}\BibitemShut {NoStop}%
\bibitem [{\citenamefont {Watanabe}\ and\ \citenamefont {Yanase}(2021)}]{watanabe2021chiral}%
  \BibitemOpen
  \bibfield  {author} {\bibinfo {author} {\bibfnamefont {Hikaru}\ \bibnamefont {Watanabe}}\ and\ \bibinfo {author} {\bibfnamefont {Youichi}\ \bibnamefont {Yanase}},\ }\bibfield  {title} {\enquote {\bibinfo {title} {{Chiral} photocurrent in parity-violating magnet and enhanced response in topological antiferromagnet},}\ }\href {\doibase 10.1103/PhysRevX.11.011001} {\bibfield  {journal} {\bibinfo  {journal} {Physical Review X}\ }\textbf {\bibinfo {volume} {11}},\ \bibinfo {pages} {011001} (\bibinfo {year} {2021})}\BibitemShut {NoStop}%
\bibitem [{\citenamefont {Kitamura}\ \emph {et~al.}(2022)\citenamefont {Kitamura}, \citenamefont {Daido},\ and\ \citenamefont {Yanase}}]{kitamura2022quantum}%
  \BibitemOpen
  \bibfield  {author} {\bibinfo {author} {\bibfnamefont {Taisei}\ \bibnamefont {Kitamura}}, \bibinfo {author} {\bibfnamefont {Akito}\ \bibnamefont {Daido}}, \ and\ \bibinfo {author} {\bibfnamefont {Youichi}\ \bibnamefont {Yanase}},\ }\bibfield  {title} {\enquote {\bibinfo {title} {{Quantum} geometric effect on {Fulde-Ferrell-Larkin-Ovchinnikov} superconductivity},}\ }\href {\doibase 10.1103/PhysRevB.106.184507} {\bibfield  {journal} {\bibinfo  {journal} {Physical Review B}\ }\textbf {\bibinfo {volume} {106}},\ \bibinfo {pages} {184507} (\bibinfo {year} {2022})}\BibitemShut {NoStop}%
\bibitem [{\citenamefont {Tasaki}(2020)}]{tasaki2020physics}%
  \BibitemOpen
  \bibfield  {author} {\bibinfo {author} {\bibfnamefont {Hal}\ \bibnamefont {Tasaki}},\ }\href {\doibase 10.1007/978-3-030-41265-4} {\emph {\bibinfo {title} {{Physics} and mathematics of quantum many-body systems}}},\ Vol.~\bibinfo {volume} {66}\ (\bibinfo  {publisher} {Springer},\ \bibinfo {year} {2020})\BibitemShut {NoStop}%
\bibitem [{sup()}]{supple}%
  \BibitemOpen
  \bibinfo {note} {For additional information, see the Supplemental Materials.}\BibitemShut {Stop}%
\bibitem [{\citenamefont {Kusakabe}\ and\ \citenamefont {Aoki}(1994)}]{kusakabe1994ferromagnetic}%
  \BibitemOpen
  \bibfield  {author} {\bibinfo {author} {\bibfnamefont {K}~\bibnamefont {Kusakabe}}\ and\ \bibinfo {author} {\bibfnamefont {H}~\bibnamefont {Aoki}},\ }\bibfield  {title} {\enquote {\bibinfo {title} {{Ferromagnetic} spin-wave theory in the multiband {Hubbard} model having a flat band},}\ }\href {\doibase 10.1103/PhysRevLett.72.144} {\bibfield  {journal} {\bibinfo  {journal} {Physical review letters}\ }\textbf {\bibinfo {volume} {72}},\ \bibinfo {pages} {144} (\bibinfo {year} {1994})}\BibitemShut {NoStop}%
\bibitem [{\citenamefont {S{\'o}lyom}(2010)}]{solyom2010fundamentals}%
  \BibitemOpen
  \bibfield  {author} {\bibinfo {author} {\bibfnamefont {Jen{\"o}}\ \bibnamefont {S{\'o}lyom}},\ }\href {\doibase 10.1007/978-3-642-04518-9} {\emph {\bibinfo {title} {{Fundamentals} of the {Physics} of {Solids:} {Volume} 3-Normal, {Broken-Symmetry,} and {Correlated} {Systems}}}},\ Vol.~\bibinfo {volume} {3}\ (\bibinfo  {publisher} {Springer Science \& Business Media},\ \bibinfo {year} {2010})\BibitemShut {NoStop}%
\bibitem [{\citenamefont {Shankar}(2012)}]{shankar2012principles}%
  \BibitemOpen
  \bibfield  {author} {\bibinfo {author} {\bibfnamefont {Ramamurti}\ \bibnamefont {Shankar}},\ }\href {\doibase 10.1007/978-1-4757-0576-8} {\emph {\bibinfo {title} {Principles of quantum mechanics}}}\ (\bibinfo  {publisher} {Springer Science \& Business Media},\ \bibinfo {year} {2012})\BibitemShut {NoStop}%
\bibitem [{\citenamefont {Griffiths}\ and\ \citenamefont {Schroeter}(2018)}]{griffiths2018introduction}%
  \BibitemOpen
  \bibfield  {author} {\bibinfo {author} {\bibfnamefont {David~J}\ \bibnamefont {Griffiths}}\ and\ \bibinfo {author} {\bibfnamefont {Darrell~F}\ \bibnamefont {Schroeter}},\ }\href {\doibase 10.1017/9781316995433} {\emph {\bibinfo {title} {Introduction to quantum mechanics}}}\ (\bibinfo  {publisher} {Cambridge university press},\ \bibinfo {year} {2018})\BibitemShut {NoStop}%
\bibitem [{\citenamefont {Wick}(1950)}]{wick1950evaluation}%
  \BibitemOpen
  \bibfield  {author} {\bibinfo {author} {\bibfnamefont {Gian-Carlo}\ \bibnamefont {Wick}},\ }\bibfield  {title} {\enquote {\bibinfo {title} {The evaluation of the collision matrix},}\ }\href {\doibase 10.1103/PhysRev.80.268} {\bibfield  {journal} {\bibinfo  {journal} {Physical review}\ }\textbf {\bibinfo {volume} {80}},\ \bibinfo {pages} {268} (\bibinfo {year} {1950})}\BibitemShut {NoStop}%
\bibitem [{\citenamefont {Giuliani}\ and\ \citenamefont {Vignale}(2008)}]{giuliani2008quantum}%
  \BibitemOpen
  \bibfield  {author} {\bibinfo {author} {\bibfnamefont {Gabriele}\ \bibnamefont {Giuliani}}\ and\ \bibinfo {author} {\bibfnamefont {Giovanni}\ \bibnamefont {Vignale}},\ }\href {\doibase 10.1017/CBO9780511619915} {\emph {\bibinfo {title} {Quantum theory of the electron liquid}}}\ (\bibinfo  {publisher} {Cambridge university press},\ \bibinfo {year} {2008})\BibitemShut {NoStop}%
\bibitem [{hal()}]{half}%
  \BibitemOpen
  \bibinfo {note} {To be precise, the kagome lattice exhibits saturated ferromagnetism when exactly half the zero-energy eigenstates including the gapless point are occupied.}\BibitemShut {Stop}%
\bibitem [{\citenamefont {Alavirad}\ and\ \citenamefont {Sau}(2020)}]{alavirad2020ferromagnetism}%
  \BibitemOpen
  \bibfield  {author} {\bibinfo {author} {\bibfnamefont {Yahya}\ \bibnamefont {Alavirad}}\ and\ \bibinfo {author} {\bibfnamefont {Jay}\ \bibnamefont {Sau}},\ }\bibfield  {title} {\enquote {\bibinfo {title} {{Ferromagnetism} and its stability from the one-magnon spectrum in twisted bilayer graphene},}\ }\href {\doibase 10.1103/PhysRevB.102.235123} {\bibfield  {journal} {\bibinfo  {journal} {Physical Review B}\ }\textbf {\bibinfo {volume} {102}},\ \bibinfo {pages} {235123} (\bibinfo {year} {2020})}\BibitemShut {NoStop}%
\bibitem [{\citenamefont {Souza}\ \emph {et~al.}(2001)\citenamefont {Souza}, \citenamefont {Marzari},\ and\ \citenamefont {Vanderbilt}}]{souza2001maximally}%
  \BibitemOpen
  \bibfield  {author} {\bibinfo {author} {\bibfnamefont {Ivo}\ \bibnamefont {Souza}}, \bibinfo {author} {\bibfnamefont {Nicola}\ \bibnamefont {Marzari}}, \ and\ \bibinfo {author} {\bibfnamefont {David}\ \bibnamefont {Vanderbilt}},\ }\bibfield  {title} {\enquote {\bibinfo {title} {{Maximally} localized {Wannier} functions for entangled energy bands},}\ }\href {\doibase 10.1103/PhysRevB.65.035109} {\bibfield  {journal} {\bibinfo  {journal} {Physical Review B}\ }\textbf {\bibinfo {volume} {65}},\ \bibinfo {pages} {035109} (\bibinfo {year} {2001})}\BibitemShut {NoStop}%
\bibitem [{\citenamefont {Huhtinen}\ \emph {et~al.}(2022)\citenamefont {Huhtinen}, \citenamefont {Herzog-Arbeitman}, \citenamefont {Chew}, \citenamefont {Bernevig},\ and\ \citenamefont {T{\"o}rm{\"a}}}]{huhtinen2022revisiting}%
  \BibitemOpen
  \bibfield  {author} {\bibinfo {author} {\bibfnamefont {Kukka-Emilia}\ \bibnamefont {Huhtinen}}, \bibinfo {author} {\bibfnamefont {Jonah}\ \bibnamefont {Herzog-Arbeitman}}, \bibinfo {author} {\bibfnamefont {Aaron}\ \bibnamefont {Chew}}, \bibinfo {author} {\bibfnamefont {Bogdan~A}\ \bibnamefont {Bernevig}}, \ and\ \bibinfo {author} {\bibfnamefont {P{\"a}ivi}\ \bibnamefont {T{\"o}rm{\"a}}},\ }\bibfield  {title} {\enquote {\bibinfo {title} {{Revisiting} flat band superconductivity: {Dependence} on minimal quantum metric and band touchings},}\ }\href {\doibase 10.1103/PhysRevB.106.014518} {\bibfield  {journal} {\bibinfo  {journal} {Physical Review B}\ }\textbf {\bibinfo {volume} {106}},\ \bibinfo {pages} {014518} (\bibinfo {year} {2022})}\BibitemShut {NoStop}%
\bibitem [{\citenamefont {Cheng}(2010)}]{cheng2010quantum}%
  \BibitemOpen
  \bibfield  {author} {\bibinfo {author} {\bibfnamefont {Ran}\ \bibnamefont {Cheng}},\ }\bibfield  {title} {\enquote {\bibinfo {title} {Quantum geometric tensor (fubini-study metric) in simple quantum system: A pedagogical introduction},}\ }\href {\doibase 10.48550/arXiv.1012.1337} {\bibfield  {journal} {\bibinfo  {journal} {arXiv preprint arXiv:1012.1337}\ } (\bibinfo {year} {2010}),\ 10.48550/arXiv.1012.1337}\BibitemShut {NoStop}%
\bibitem [{\citenamefont {Provost}\ and\ \citenamefont {Vallee}(1980)}]{provost1980riemannian}%
  \BibitemOpen
  \bibfield  {author} {\bibinfo {author} {\bibfnamefont {JP}~\bibnamefont {Provost}}\ and\ \bibinfo {author} {\bibfnamefont {G}~\bibnamefont {Vallee}},\ }\bibfield  {title} {\enquote {\bibinfo {title} {{Riemannian} structure on manifolds of quantum states},}\ }\href {\doibase 10.1007/BF02193559} {\bibfield  {journal} {\bibinfo  {journal} {Communications in Mathematical Physics}\ }\textbf {\bibinfo {volume} {76}},\ \bibinfo {pages} {289--301} (\bibinfo {year} {1980})}\BibitemShut {NoStop}%
\bibitem [{\citenamefont {Resta}(2011)}]{resta2011insulating}%
  \BibitemOpen
  \bibfield  {author} {\bibinfo {author} {\bibfnamefont {Raffaele}\ \bibnamefont {Resta}},\ }\bibfield  {title} {\enquote {\bibinfo {title} {{The} insulating state of matter: a geometrical theory},}\ }\href {\doibase 10.1140/epjb/e2010-10874-4} {\bibfield  {journal} {\bibinfo  {journal} {The European Physical Journal B}\ }\textbf {\bibinfo {volume} {79}},\ \bibinfo {pages} {121--137} (\bibinfo {year} {2011})}\BibitemShut {NoStop}%
\bibitem [{\citenamefont {Berry}(1984)}]{berry1984quantal}%
  \BibitemOpen
  \bibfield  {author} {\bibinfo {author} {\bibfnamefont {Michael~Victor}\ \bibnamefont {Berry}},\ }\bibfield  {title} {\enquote {\bibinfo {title} {Quantal phase factors accompanying adiabatic changes},}\ }\href {\doibase 10.1098/rspa.1984.0023} {\bibfield  {journal} {\bibinfo  {journal} {Proceedings of the Royal Society of London. A. Mathematical and Physical Sciences}\ }\textbf {\bibinfo {volume} {392}},\ \bibinfo {pages} {45--57} (\bibinfo {year} {1984})}\BibitemShut {NoStop}%
\bibitem [{\citenamefont {Hwang}\ \emph {et~al.}(2021{\natexlab{b}})\citenamefont {Hwang}, \citenamefont {Jung}, \citenamefont {Rhim},\ and\ \citenamefont {Yang}}]{hwang2021wave}%
  \BibitemOpen
  \bibfield  {author} {\bibinfo {author} {\bibfnamefont {Yoonseok}\ \bibnamefont {Hwang}}, \bibinfo {author} {\bibfnamefont {Junseo}\ \bibnamefont {Jung}}, \bibinfo {author} {\bibfnamefont {Jun-Won}\ \bibnamefont {Rhim}}, \ and\ \bibinfo {author} {\bibfnamefont {Bohm-Jung}\ \bibnamefont {Yang}},\ }\bibfield  {title} {\enquote {\bibinfo {title} {{Wave-function} geometry of band crossing points in two dimensions},}\ }\href {\doibase 10.1103/PhysRevB.103.L241102} {\bibfield  {journal} {\bibinfo  {journal} {Physical Review B}\ }\textbf {\bibinfo {volume} {103}},\ \bibinfo {pages} {L241102} (\bibinfo {year} {2021}{\natexlab{b}})}\BibitemShut {NoStop}%
\bibitem [{\citenamefont {Repellin}\ \emph {et~al.}(2020)\citenamefont {Repellin}, \citenamefont {Dong}, \citenamefont {Zhang},\ and\ \citenamefont {Senthil}}]{repellin2020ferromagnetism}%
  \BibitemOpen
  \bibfield  {author} {\bibinfo {author} {\bibfnamefont {C{\'e}cile}\ \bibnamefont {Repellin}}, \bibinfo {author} {\bibfnamefont {Zhihuan}\ \bibnamefont {Dong}}, \bibinfo {author} {\bibfnamefont {Ya-Hui}\ \bibnamefont {Zhang}}, \ and\ \bibinfo {author} {\bibfnamefont {T}~\bibnamefont {Senthil}},\ }\bibfield  {title} {\enquote {\bibinfo {title} {{Ferromagnetism} in narrow bands of moir{\'e} superlattices},}\ }\href {\doibase 10.1103/PhysRevLett.124.187601} {\bibfield  {journal} {\bibinfo  {journal} {Physical Review Letters}\ }\textbf {\bibinfo {volume} {124}},\ \bibinfo {pages} {187601} (\bibinfo {year} {2020})}\BibitemShut {NoStop}%
\bibitem [{\citenamefont {Marzari}\ and\ \citenamefont {Vanderbilt}(1997)}]{marzari1997maximally}%
  \BibitemOpen
  \bibfield  {author} {\bibinfo {author} {\bibfnamefont {Nicola}\ \bibnamefont {Marzari}}\ and\ \bibinfo {author} {\bibfnamefont {David}\ \bibnamefont {Vanderbilt}},\ }\bibfield  {title} {\enquote {\bibinfo {title} {{Maximally} localized generalized {Wannier} functions for composite energy bands},}\ }\href {\doibase 10.1103/PhysRevB.56.12847} {\bibfield  {journal} {\bibinfo  {journal} {Physical review B}\ }\textbf {\bibinfo {volume} {56}},\ \bibinfo {pages} {12847} (\bibinfo {year} {1997})}\BibitemShut {NoStop}%
\bibitem [{\citenamefont {Lieb}(1989)}]{lieb1989two}%
  \BibitemOpen
  \bibfield  {author} {\bibinfo {author} {\bibfnamefont {Elliott~H}\ \bibnamefont {Lieb}},\ }\bibfield  {title} {\enquote {\bibinfo {title} {{Two} theorems on the {Hubbard} model},}\ }\href {\doibase 10.1103/PhysRevLett.62.1201} {\bibfield  {journal} {\bibinfo  {journal} {Physical review letters}\ }\textbf {\bibinfo {volume} {62}},\ \bibinfo {pages} {1201} (\bibinfo {year} {1989})}\BibitemShut {NoStop}%
\bibitem [{\citenamefont {Fazekas}\ \emph {et~al.}(1990)\citenamefont {Fazekas}, \citenamefont {Menge},\ and\ \citenamefont {M{\"u}ller-Hartmann}}]{fazekas1990ground}%
  \BibitemOpen
  \bibfield  {author} {\bibinfo {author} {\bibfnamefont {P}~\bibnamefont {Fazekas}}, \bibinfo {author} {\bibfnamefont {B}~\bibnamefont {Menge}}, \ and\ \bibinfo {author} {\bibfnamefont {E}~\bibnamefont {M{\"u}ller-Hartmann}},\ }\bibfield  {title} {\enquote {\bibinfo {title} {{Ground} state phase diagram of the infinite dimensional {Hubbard} model: {A} variational study},}\ }\href {\doibase 10.1007/BF01317359} {\bibfield  {journal} {\bibinfo  {journal} {Zeitschrift f{\"u}r Physik B Condensed Matter}\ }\textbf {\bibinfo {volume} {78}},\ \bibinfo {pages} {69--80} (\bibinfo {year} {1990})}\BibitemShut {NoStop}%
\bibitem [{\citenamefont {Lee}\ and\ \citenamefont {Lee}(2005)}]{lee2005u}%
  \BibitemOpen
  \bibfield  {author} {\bibinfo {author} {\bibfnamefont {Sung-Sik}\ \bibnamefont {Lee}}\ and\ \bibinfo {author} {\bibfnamefont {Patrick~A.}\ \bibnamefont {Lee}},\ }\bibfield  {title} {\enquote {\bibinfo {title} {{U(1)} {Gauge} {Theory} of the {Hubbard} {Model:} {Spin} {Liquid} {States} and {Possible} {Application} to {$\kappa-\text{(BEDT}-\text{TTF)}_2 \text{Cu}_2 \text{(CN)}_3$}},}\ }\href {\doibase 10.1103/PhysRevLett.95.036403} {\bibfield  {journal} {\bibinfo  {journal} {Phys. Rev. Lett.}\ }\textbf {\bibinfo {volume} {95}},\ \bibinfo {pages} {036403} (\bibinfo {year} {2005})}\BibitemShut {NoStop}%
\bibitem [{\citenamefont {Szasz}\ \emph {et~al.}(2020)\citenamefont {Szasz}, \citenamefont {Motruk}, \citenamefont {Zaletel},\ and\ \citenamefont {Moore}}]{szasz2020chiral}%
  \BibitemOpen
  \bibfield  {author} {\bibinfo {author} {\bibfnamefont {Aaron}\ \bibnamefont {Szasz}}, \bibinfo {author} {\bibfnamefont {Johannes}\ \bibnamefont {Motruk}}, \bibinfo {author} {\bibfnamefont {Michael~P.}\ \bibnamefont {Zaletel}}, \ and\ \bibinfo {author} {\bibfnamefont {Joel~E.}\ \bibnamefont {Moore}},\ }\bibfield  {title} {\enquote {\bibinfo {title} {{Chiral} {Spin} {Liquid} {Phase} of the {Triangular} {Lattice} {Hubbard} {Model:} {A} {Density} {Matrix} {Renormalization} {Group} {Study}},}\ }\href {\doibase 10.1103/PhysRevX.10.021042} {\bibfield  {journal} {\bibinfo  {journal} {Phys. Rev. X}\ }\textbf {\bibinfo {volume} {10}},\ \bibinfo {pages} {021042} (\bibinfo {year} {2020})}\BibitemShut {NoStop}%
\bibitem [{\citenamefont {Herzog-Arbeitman}\ \emph {et~al.}(2022{\natexlab{b}})\citenamefont {Herzog-Arbeitman}, \citenamefont {Peri}, \citenamefont {Schindler}, \citenamefont {Huber},\ and\ \citenamefont {Bernevig}}]{herzog2022superfluid}%
  \BibitemOpen
  \bibfield  {author} {\bibinfo {author} {\bibfnamefont {Jonah}\ \bibnamefont {Herzog-Arbeitman}}, \bibinfo {author} {\bibfnamefont {Valerio}\ \bibnamefont {Peri}}, \bibinfo {author} {\bibfnamefont {Frank}\ \bibnamefont {Schindler}}, \bibinfo {author} {\bibfnamefont {Sebastian~D}\ \bibnamefont {Huber}}, \ and\ \bibinfo {author} {\bibfnamefont {B~Andrei}\ \bibnamefont {Bernevig}},\ }\bibfield  {title} {\enquote {\bibinfo {title} {{Superfluid} weight bounds from symmetry and quantum geometry in flat bands},}\ }\href {\doibase 10.1103/PhysRevLett.128.087002} {\bibfield  {journal} {\bibinfo  {journal} {Physical review letters}\ }\textbf {\bibinfo {volume} {128}},\ \bibinfo {pages} {087002} (\bibinfo {year} {2022}{\natexlab{b}})}\BibitemShut {NoStop}%
\bibitem [{\citenamefont {McGuire}\ \emph {et~al.}(2015)\citenamefont {McGuire}, \citenamefont {Dixit}, \citenamefont {Cooper},\ and\ \citenamefont {Sales}}]{mcguire2015coupling}%
  \BibitemOpen
  \bibfield  {author} {\bibinfo {author} {\bibfnamefont {Michael~A}\ \bibnamefont {McGuire}}, \bibinfo {author} {\bibfnamefont {Hemant}\ \bibnamefont {Dixit}}, \bibinfo {author} {\bibfnamefont {Valentino~R}\ \bibnamefont {Cooper}}, \ and\ \bibinfo {author} {\bibfnamefont {Brian~C}\ \bibnamefont {Sales}},\ }\bibfield  {title} {\enquote {\bibinfo {title} {{Coupling} of crystal structure and magnetism in the layered, ferromagnetic insulator {CrI$_3$}},}\ }\href {\doibase 10.1021/cm504242t} {\bibfield  {journal} {\bibinfo  {journal} {Chemistry of Materials}\ }\textbf {\bibinfo {volume} {27}},\ \bibinfo {pages} {612--620} (\bibinfo {year} {2015})}\BibitemShut {NoStop}%
\bibitem [{\citenamefont {Song}\ \emph {et~al.}(2006)\citenamefont {Song}, \citenamefont {Geng}, \citenamefont {Zeng}, \citenamefont {Wang}, \citenamefont {Shen}, \citenamefont {Pan}, \citenamefont {Xie}, \citenamefont {Liu}, \citenamefont {Zhou},\ and\ \citenamefont {Fan}}]{song2006giant}%
  \BibitemOpen
  \bibfield  {author} {\bibinfo {author} {\bibfnamefont {C}~\bibnamefont {Song}}, \bibinfo {author} {\bibfnamefont {KW}~\bibnamefont {Geng}}, \bibinfo {author} {\bibfnamefont {F}~\bibnamefont {Zeng}}, \bibinfo {author} {\bibfnamefont {XB}~\bibnamefont {Wang}}, \bibinfo {author} {\bibfnamefont {YX}~\bibnamefont {Shen}}, \bibinfo {author} {\bibfnamefont {F}~\bibnamefont {Pan}}, \bibinfo {author} {\bibfnamefont {YN}~\bibnamefont {Xie}}, \bibinfo {author} {\bibfnamefont {T}~\bibnamefont {Liu}}, \bibinfo {author} {\bibfnamefont {HT}~\bibnamefont {Zhou}}, \ and\ \bibinfo {author} {\bibfnamefont {Z}~\bibnamefont {Fan}},\ }\bibfield  {title} {\enquote {\bibinfo {title} {{Giant} magnetic moment in an anomalous ferromagnetic insulator: {Co-doped} {ZnO}},}\ }\href {\doibase 10.1103/PhysRevB.73.024405} {\bibfield  {journal} {\bibinfo  {journal} {Physical Review B}\ }\textbf {\bibinfo {volume} {73}},\ \bibinfo {pages} {024405} (\bibinfo {year} {2006})}\BibitemShut {NoStop}%
\bibitem [{\citenamefont {Meng}\ \emph {et~al.}(2018)\citenamefont {Meng}, \citenamefont {Guo}, \citenamefont {Cui}, \citenamefont {Ma}, \citenamefont {Zhao}, \citenamefont {Lu}, \citenamefont {Xu}, \citenamefont {Wang}, \citenamefont {Hu}, \citenamefont {Fu} \emph {et~al.}}]{meng2018strain}%
  \BibitemOpen
  \bibfield  {author} {\bibinfo {author} {\bibfnamefont {Dechao}\ \bibnamefont {Meng}}, \bibinfo {author} {\bibfnamefont {Hongli}\ \bibnamefont {Guo}}, \bibinfo {author} {\bibfnamefont {Zhangzhang}\ \bibnamefont {Cui}}, \bibinfo {author} {\bibfnamefont {Chao}\ \bibnamefont {Ma}}, \bibinfo {author} {\bibfnamefont {Jin}\ \bibnamefont {Zhao}}, \bibinfo {author} {\bibfnamefont {Jiangbo}\ \bibnamefont {Lu}}, \bibinfo {author} {\bibfnamefont {Hui}\ \bibnamefont {Xu}}, \bibinfo {author} {\bibfnamefont {Zhicheng}\ \bibnamefont {Wang}}, \bibinfo {author} {\bibfnamefont {Xiang}\ \bibnamefont {Hu}}, \bibinfo {author} {\bibfnamefont {Zhengping}\ \bibnamefont {Fu}},  \emph {et~al.},\ }\bibfield  {title} {\enquote {\bibinfo {title} {{Strain-induced} high-temperature perovskite ferromagnetic insulator},}\ }\href {\doibase 10.1073/pnas.1707817115} {\bibfield  {journal} {\bibinfo  {journal} {Proceedings of the National Academy of Sciences}\ }\textbf {\bibinfo {volume} {115}},\ \bibinfo {pages} {2873--2877} (\bibinfo {year}
  {2018})}\BibitemShut {NoStop}%
\bibitem [{\citenamefont {Zhang}\ \emph {et~al.}(2016)\citenamefont {Zhang}, \citenamefont {Zhao}, \citenamefont {Song}, \citenamefont {Jia}, \citenamefont {Shi},\ and\ \citenamefont {Han}}]{zhang2016magnetic}%
  \BibitemOpen
  \bibfield  {author} {\bibinfo {author} {\bibfnamefont {Xiao}\ \bibnamefont {Zhang}}, \bibinfo {author} {\bibfnamefont {Yuelei}\ \bibnamefont {Zhao}}, \bibinfo {author} {\bibfnamefont {Qi}~\bibnamefont {Song}}, \bibinfo {author} {\bibfnamefont {Shuang}\ \bibnamefont {Jia}}, \bibinfo {author} {\bibfnamefont {Jing}\ \bibnamefont {Shi}}, \ and\ \bibinfo {author} {\bibfnamefont {Wei}\ \bibnamefont {Han}},\ }\bibfield  {title} {\enquote {\bibinfo {title} {{Magnetic} anisotropy of the single-crystalline ferromagnetic insulator {Cr$_2$Ge$_2$Te$_6$}},}\ }\href {\doibase 10.7567/JJAP.55.033001} {\bibfield  {journal} {\bibinfo  {journal} {Japanese Journal of Applied Physics}\ }\textbf {\bibinfo {volume} {55}},\ \bibinfo {pages} {033001} (\bibinfo {year} {2016})}\BibitemShut {NoStop}%
\bibitem [{\citenamefont {Liu}\ \emph {et~al.}(2018)\citenamefont {Liu}, \citenamefont {Sun}, \citenamefont {Kumar}, \citenamefont {Muechler}, \citenamefont {Sun}, \citenamefont {Jiao}, \citenamefont {Yang}, \citenamefont {Liu}, \citenamefont {Liang}, \citenamefont {Xu} \emph {et~al.}}]{liu2018giant}%
  \BibitemOpen
  \bibfield  {author} {\bibinfo {author} {\bibfnamefont {Enke}\ \bibnamefont {Liu}}, \bibinfo {author} {\bibfnamefont {Yan}\ \bibnamefont {Sun}}, \bibinfo {author} {\bibfnamefont {Nitesh}\ \bibnamefont {Kumar}}, \bibinfo {author} {\bibfnamefont {Lukas}\ \bibnamefont {Muechler}}, \bibinfo {author} {\bibfnamefont {Aili}\ \bibnamefont {Sun}}, \bibinfo {author} {\bibfnamefont {Lin}\ \bibnamefont {Jiao}}, \bibinfo {author} {\bibfnamefont {Shuo-Ying}\ \bibnamefont {Yang}}, \bibinfo {author} {\bibfnamefont {Defa}\ \bibnamefont {Liu}}, \bibinfo {author} {\bibfnamefont {Aiji}\ \bibnamefont {Liang}}, \bibinfo {author} {\bibfnamefont {Qiunan}\ \bibnamefont {Xu}},  \emph {et~al.},\ }\bibfield  {title} {\enquote {\bibinfo {title} {{Giant} anomalous {Hall} effect in a ferromagnetic kagome-lattice semimetal},}\ }\href {\doibase 10.1038/s41567-018-0234-5} {\bibfield  {journal} {\bibinfo  {journal} {Nature physics}\ }\textbf {\bibinfo {volume} {14}},\ \bibinfo {pages} {1125--1131} (\bibinfo {year} {2018})}\BibitemShut
  {NoStop}%
\bibitem [{\citenamefont {Kim}\ \emph {et~al.}(2018)\citenamefont {Kim}, \citenamefont {Seo}, \citenamefont {Lee}, \citenamefont {Ko}, \citenamefont {Kim}, \citenamefont {Jang}, \citenamefont {Ok}, \citenamefont {Lee}, \citenamefont {Jo}, \citenamefont {Kang} \emph {et~al.}}]{kim2018large}%
  \BibitemOpen
  \bibfield  {author} {\bibinfo {author} {\bibfnamefont {Kyoo}\ \bibnamefont {Kim}}, \bibinfo {author} {\bibfnamefont {Junho}\ \bibnamefont {Seo}}, \bibinfo {author} {\bibfnamefont {Eunwoo}\ \bibnamefont {Lee}}, \bibinfo {author} {\bibfnamefont {K-T}\ \bibnamefont {Ko}}, \bibinfo {author} {\bibfnamefont {BS}~\bibnamefont {Kim}}, \bibinfo {author} {\bibfnamefont {Bo~Gyu}\ \bibnamefont {Jang}}, \bibinfo {author} {\bibfnamefont {Jong~Mok}\ \bibnamefont {Ok}}, \bibinfo {author} {\bibfnamefont {Jinwon}\ \bibnamefont {Lee}}, \bibinfo {author} {\bibfnamefont {Youn~Jung}\ \bibnamefont {Jo}}, \bibinfo {author} {\bibfnamefont {Woun}\ \bibnamefont {Kang}},  \emph {et~al.},\ }\bibfield  {title} {\enquote {\bibinfo {title} {{Large} anomalous {Hall} current induced by topological nodal lines in a ferromagnetic van der {Waals} semimetal},}\ }\href {\doibase 10.1038/s41563-018-0132-3} {\bibfield  {journal} {\bibinfo  {journal} {Nature materials}\ }\textbf {\bibinfo {volume} {17}},\ \bibinfo {pages} {794--799} (\bibinfo
  {year} {2018})}\BibitemShut {NoStop}%
\bibitem [{\citenamefont {Jin}\ \emph {et~al.}(2017)\citenamefont {Jin}, \citenamefont {Wang}, \citenamefont {Chen}, \citenamefont {Zhao}, \citenamefont {Zhao},\ and\ \citenamefont {Xu}}]{jin2017ferromagnetic}%
  \BibitemOpen
  \bibfield  {author} {\bibinfo {author} {\bibfnamefont {YJ}~\bibnamefont {Jin}}, \bibinfo {author} {\bibfnamefont {R}~\bibnamefont {Wang}}, \bibinfo {author} {\bibfnamefont {ZJ}~\bibnamefont {Chen}}, \bibinfo {author} {\bibfnamefont {JZ}~\bibnamefont {Zhao}}, \bibinfo {author} {\bibfnamefont {YJ}~\bibnamefont {Zhao}}, \ and\ \bibinfo {author} {\bibfnamefont {H}~\bibnamefont {Xu}},\ }\bibfield  {title} {\enquote {\bibinfo {title} {{Ferromagnetic} {Weyl} semimetal phase in a tetragonal structure},}\ }\href {\doibase 10.1103/PhysRevB.96.201102} {\bibfield  {journal} {\bibinfo  {journal} {Physical Review B}\ }\textbf {\bibinfo {volume} {96}},\ \bibinfo {pages} {201102} (\bibinfo {year} {2017})}\BibitemShut {NoStop}%
\bibitem [{\citenamefont {Liu}\ \emph {et~al.}(2019)\citenamefont {Liu}, \citenamefont {Liang}, \citenamefont {Liu}, \citenamefont {Xu}, \citenamefont {Li}, \citenamefont {Chen}, \citenamefont {Pei}, \citenamefont {Shi}, \citenamefont {Mo}, \citenamefont {Dudin} \emph {et~al.}}]{liu2019magnetic}%
  \BibitemOpen
  \bibfield  {author} {\bibinfo {author} {\bibfnamefont {DF}~\bibnamefont {Liu}}, \bibinfo {author} {\bibfnamefont {AJ}~\bibnamefont {Liang}}, \bibinfo {author} {\bibfnamefont {EK}~\bibnamefont {Liu}}, \bibinfo {author} {\bibfnamefont {QN}~\bibnamefont {Xu}}, \bibinfo {author} {\bibfnamefont {YW}~\bibnamefont {Li}}, \bibinfo {author} {\bibfnamefont {C}~\bibnamefont {Chen}}, \bibinfo {author} {\bibfnamefont {D}~\bibnamefont {Pei}}, \bibinfo {author} {\bibfnamefont {WJ}~\bibnamefont {Shi}}, \bibinfo {author} {\bibfnamefont {SK}~\bibnamefont {Mo}}, \bibinfo {author} {\bibfnamefont {P}~\bibnamefont {Dudin}},  \emph {et~al.},\ }\bibfield  {title} {\enquote {\bibinfo {title} {{Magnetic} {Weyl} semimetal phase in a {Kagom{\'e}} crystal},}\ }\href {\doibase 10.1126/science.aav2873} {\bibfield  {journal} {\bibinfo  {journal} {Science}\ }\textbf {\bibinfo {volume} {365}},\ \bibinfo {pages} {1282--1285} (\bibinfo {year} {2019})}\BibitemShut {NoStop}%
\bibitem [{\citenamefont {Song}\ \emph {et~al.}(2020)\citenamefont {Song}, \citenamefont {Elcoro},\ and\ \citenamefont {Bernevig}}]{song2020twisted}%
  \BibitemOpen
  \bibfield  {author} {\bibinfo {author} {\bibfnamefont {Zhi-Da}\ \bibnamefont {Song}}, \bibinfo {author} {\bibfnamefont {Luis}\ \bibnamefont {Elcoro}}, \ and\ \bibinfo {author} {\bibfnamefont {B~Andrei}\ \bibnamefont {Bernevig}},\ }\bibfield  {title} {\enquote {\bibinfo {title} {{Twisted} bulk-boundary correspondence of fragile topology},}\ }\href {\doibase 10.1126/science.aaz7650} {\bibfield  {journal} {\bibinfo  {journal} {Science}\ }\textbf {\bibinfo {volume} {367}},\ \bibinfo {pages} {794--797} (\bibinfo {year} {2020})}\BibitemShut {NoStop}%
\bibitem [{\citenamefont {Jung}\ \emph {et~al.}(2011)\citenamefont {Jung}, \citenamefont {Zhang},\ and\ \citenamefont {MacDonald}}]{jung2011lattice}%
  \BibitemOpen
  \bibfield  {author} {\bibinfo {author} {\bibfnamefont {Jeil}\ \bibnamefont {Jung}}, \bibinfo {author} {\bibfnamefont {Fan}\ \bibnamefont {Zhang}}, \ and\ \bibinfo {author} {\bibfnamefont {Allan~H}\ \bibnamefont {MacDonald}},\ }\bibfield  {title} {\enquote {\bibinfo {title} {{Lattice} theory of pseudospin ferromagnetism in bilayer graphene: {Competing} interaction-induced quantum {Hall} states},}\ }\href {\doibase 10.1103/PhysRevB.83.115408} {\bibfield  {journal} {\bibinfo  {journal} {Physical Review B—Condensed Matter and Materials Physics}\ }\textbf {\bibinfo {volume} {83}},\ \bibinfo {pages} {115408} (\bibinfo {year} {2011})}\BibitemShut {NoStop}%
\bibitem [{\citenamefont {L{\"o}wdin}(1950)}]{lowdin1950non}%
  \BibitemOpen
  \bibfield  {author} {\bibinfo {author} {\bibfnamefont {Per-Olov}\ \bibnamefont {L{\"o}wdin}},\ }\bibfield  {title} {\enquote {\bibinfo {title} {{On} the non-orthogonality problem connected with the use of atomic wave functions in the theory of molecules and crystals},}\ }\href {\doibase 10.1063/1.1747632} {\bibfield  {journal} {\bibinfo  {journal} {The Journal of Chemical Physics}\ }\textbf {\bibinfo {volume} {18}},\ \bibinfo {pages} {365--375} (\bibinfo {year} {1950})}\BibitemShut {NoStop}%
\bibitem [{\citenamefont {Slater}\ and\ \citenamefont {Koster}(1954)}]{slater1954simplified}%
  \BibitemOpen
  \bibfield  {author} {\bibinfo {author} {\bibfnamefont {John~C}\ \bibnamefont {Slater}}\ and\ \bibinfo {author} {\bibfnamefont {George~F}\ \bibnamefont {Koster}},\ }\bibfield  {title} {\enquote {\bibinfo {title} {{Simplified} {LCAO} method for the periodic potential problem},}\ }\href {\doibase 10.1103/PhysRev.94.1498} {\bibfield  {journal} {\bibinfo  {journal} {Physical review}\ }\textbf {\bibinfo {volume} {94}},\ \bibinfo {pages} {1498} (\bibinfo {year} {1954})}\BibitemShut {NoStop}%
\bibitem [{\citenamefont {Goringe}\ \emph {et~al.}(1997)\citenamefont {Goringe}, \citenamefont {Bowler},\ and\ \citenamefont {Hernandez}}]{goringe1997tight}%
  \BibitemOpen
  \bibfield  {author} {\bibinfo {author} {\bibfnamefont {CM}~\bibnamefont {Goringe}}, \bibinfo {author} {\bibfnamefont {DR}~\bibnamefont {Bowler}}, \ and\ \bibinfo {author} {\bibfnamefont {E}~\bibnamefont {Hernandez}},\ }\bibfield  {title} {\enquote {\bibinfo {title} {{Tight-binding} modelling of materials},}\ }\href {\doibase 10.1088/0034-4885/60/12/001} {\bibfield  {journal} {\bibinfo  {journal} {Reports on Progress in Physics}\ }\textbf {\bibinfo {volume} {60}},\ \bibinfo {pages} {1447} (\bibinfo {year} {1997})}\BibitemShut {NoStop}%
\bibitem [{\citenamefont {Jozsa}(1994)}]{jozsa1994fidelity}%
  \BibitemOpen
  \bibfield  {author} {\bibinfo {author} {\bibfnamefont {Richard}\ \bibnamefont {Jozsa}},\ }\bibfield  {title} {\enquote {\bibinfo {title} {{Fidelity} for mixed quantum states},}\ }\href {\doibase 10.1080/09500349414552171} {\bibfield  {journal} {\bibinfo  {journal} {Journal of modern optics}\ }\textbf {\bibinfo {volume} {41}},\ \bibinfo {pages} {2315--2323} (\bibinfo {year} {1994})}\BibitemShut {NoStop}%
\bibitem [{\citenamefont {G{\"u}ttinger}(1932)}]{guttinger1932verhalten}%
  \BibitemOpen
  \bibfield  {author} {\bibinfo {author} {\bibfnamefont {Paul}\ \bibnamefont {G{\"u}ttinger}},\ }\bibfield  {title} {\enquote {\bibinfo {title} {Das verhalten von atomen im magnetischen drehfeld},}\ }\href {\doibase 10.1007/BF01351211} {\bibfield  {journal} {\bibinfo  {journal} {Zeitschrift f{\"u}r Physik}\ }\textbf {\bibinfo {volume} {73}},\ \bibinfo {pages} {169--184} (\bibinfo {year} {1932})}\BibitemShut {NoStop}%
\bibitem [{\citenamefont {Feynman}(1939)}]{feynman1939forces}%
  \BibitemOpen
  \bibfield  {author} {\bibinfo {author} {\bibfnamefont {Richard~Phillips}\ \bibnamefont {Feynman}},\ }\bibfield  {title} {\enquote {\bibinfo {title} {Forces in molecules},}\ }\href {\doibase 10.1103/PhysRev.56.340} {\bibfield  {journal} {\bibinfo  {journal} {Physical review}\ }\textbf {\bibinfo {volume} {56}},\ \bibinfo {pages} {340} (\bibinfo {year} {1939})}\BibitemShut {NoStop}%
\bibitem [{jon()}]{jonah}%
  \BibitemOpen
  \bibinfo {note} {Jonah-Herzog Arbeitman, private communication}\BibitemShut {NoStop}%
\bibitem [{\citenamefont {Haldane}(1988)}]{haldane1988model}%
  \BibitemOpen
  \bibfield  {author} {\bibinfo {author} {\bibfnamefont {F~Duncan~M}\ \bibnamefont {Haldane}},\ }\bibfield  {title} {\enquote {\bibinfo {title} {Model for a quantum {Hall} effect without {Landau} levels: {Condensed}-matter realization of the {``parity} {anomaly"}},}\ }\href {\doibase 10.1103/PhysRevLett.61.2015} {\bibfield  {journal} {\bibinfo  {journal} {Physical review letters}\ }\textbf {\bibinfo {volume} {61}},\ \bibinfo {pages} {2015} (\bibinfo {year} {1988})}\BibitemShut {NoStop}%
\bibitem [{\citenamefont {Bradlyn}\ and\ \citenamefont {Iraola}(2022)}]{bradlyn2022lecture}%
  \BibitemOpen
  \bibfield  {author} {\bibinfo {author} {\bibfnamefont {Barry}\ \bibnamefont {Bradlyn}}\ and\ \bibinfo {author} {\bibfnamefont {Mikel}\ \bibnamefont {Iraola}},\ }\bibfield  {title} {\enquote {\bibinfo {title} {Lecture notes on {Berry} phases and topology},}\ }\href {\doibase 10.21468/SciPostPhysLectNotes.51} {\bibfield  {journal} {\bibinfo  {journal} {SciPost Physics Lecture Notes}\ ,\ \bibinfo {pages} {051}} (\bibinfo {year} {2022})}\BibitemShut {NoStop}%
\end{thebibliography}%

\let\addcontentsline\oldaddcontentsline

\clearpage
\onecolumngrid
\begin{center}
\textbf{\large Supplemental Material for ``\ourtitle"}
\end{center}
\setcounter{section}{0}
\setcounter{figure}{0}
\setcounter{equation}{0}
\setcounter{table}{0}
\renewcommand{\thefigure}{S\arabic{figure}}
\renewcommand{\theequation}{S\arabic{equation}}
\renewcommand{\thesection}{S\arabic{section}}
\renewcommand{\thetable}{S\arabic{table}}
\tableofcontents
\hfill \\
\onecolumngrid


\section{Tight-binding conventions}
\label{app:Conv}

%
In this Appendix, we introduce the tight-binding notations used throughout this letter.
We consider a $d$-dimensional periodic lattice with $\nc$ unit cells, spanned by the primitive lattice vectors $\bba_1, \dots, \bba_d$.
The reciprocal lattice vectors are defined by $\bba_i \cdot \bb{b}_j = 2\pi \delta_{ij}$.
We denote the momentum as
\ba
\bk = \sum_{i} k_i \bb{b}_i = \sum_{\mu}  k_{\mu} \ehat_{\mu},
\label{seq:momentum}
\ea
where $i = 1, \dots, d$ and $\mu = x, y, \dots$ index the crystalline and orthogonal coordinates, respectively.
The first Brillouin zone is given by $k_i \in [0, 1)$.
The tight-binding Hilbert space is generated by $\cdag_{\bR \al \sg}$, which creates  L\"{o}wdin orbitals at the unit cell $\bR$, orbital $\al = 1, \dots, \norb$, and spin $\sg = \ua, \da$~\cite{lowdin1950non,slater1954simplified,goringe1997tight}.
We denote the orbital position relative to the unit cell as $\bb{r}_{\al}$.
The momentum space operators are denoted as
\ba
\cdag_{\bk \al \sg} = \frac{1}{\sqrt{\nc}} \sum_{\bR} \ee{i \bk \cdot (\bR + \bb r_{\al})} \cdag_{\bR \al \sg},
\quad
\cdag_{\bR\al\sg} = \frac{1}{\sqrt{\nc}} \sum_{\bk} e^{-i\bk \cdot (\bR + \bb r_{\al})}.
\label{seq:ck}
\ea
Note that the position dependence of the orbitals has been taken account in this convention, which is crucial when dealing with quantum geometry.
We consider the standard Hubbard model with repulsive interaction $U > 0$~\cite{hubbard1963electron,gutzwiller1963effect,kanamori1963electron}
\ba
H
&=
H_{kin} + H_{int}
\nn
&=
\sum_{\bR \bR' \al \be \sg} t_{\bR, \bR'}^{\al \be} \cdag_{\bR \al \sg} c_{\bR' \be \sg} + U \sum_{\bR \al} \cdag_{\bR \al \ua} \cdag_{\bR \al \da} c_{\bR \al \da} c_{\bR \al \ua}
\nn
&=
\sum_{\bk \al \be \sg} h(\bk)_{\al \be} \cdag_{\bk \al \sg} c_{\bk \be \sg}
+
\uc \sum_{\bk \bk' \bq \al} \cdag_{\bk + \bq \al \ua} \cdag_{\bk' - \bq \al \da} c_{\bk' \al \da} c_{\bk \al \ua},
\label{seq:Hubbard}
\ea
where we have neglected Umklapp scattering processes in the last line.
We denote the eigensystem of the single-particle Hamiltonian matrix $h(\bk)_{\al \be} = \sum_{\bR} \ee{-i \bk \cdot (\bR + \bb{r}_{\al} - \bb{r}_{\be})} t_{\bR \bb 0}^{\al \be}$ as
\ba
h(\bk) \ket{u_n (\bk)} = E_n (\bk) \ket{u_n (\bk)}
\quad
(n = 1, \dots, \norb),
\ea
where $\ket{u_n(\bk)}$ is an $\norb \times 1$ column vector.
Then, the noninteracting band electron operators are given by
\ba
\pdag_{n\bk\sg} = \sum_{\al} \ket{u_n(\bk)}_{\al} \cdag_{\bk \al \sg},
\ea
where $\ket{u_n(\bk)}_{\al}$ denotes the $\al$-th component of the corresponding vector.
We fix the electron number to $\ntot = \nc \nocc$, and assume that the saturated ferromagnetic state obtained by filling one spin sector of $H_{kin}$ has $\nocc$ occupied bands at every $\bk$, which means that the system becomes a ferromagnetic insulator or semimetal.
This does not mean that the noninteracting system with $U=0$ is insulating; in general, it is expected to be a paramagnetic metal.
Throughout this Letter, we use the term ``occupied bands" to denote the $\nocc$ lowest bands of $h(\bk)$.
The eigenvectors of the occupied bands form an $\norb \times \nocc$ matrix
\ba
\uk = (\ket{u_1(\bk)}, \dots, \ket{u_{\nocc}(\bk)}),
\ea
which satisfies
\ba
\uk \ukdag = \sum_{n} \kbr{u_n(\bk)}{u_n(\bk)} = \proj,
\quad
\ukdag \uk = \id_{\nocc \times \nocc},
\ea
where $\proj$ is the projector, and $\id_{\nocc \times \nocc}$ is the $\nocc \times \nocc$ identity matrix.
The ambiguity in choosing $\uk$ at band crossing points is avoided either by shifting the momentum space grid, or choosing the gauge which forms the maximally localized Wannier functions for entangled bands~\cite{souza2001maximally}.
%

\section{Properties of the quantum geometric tensor}
\label{app:qg}
In this appendix, we formulate the quantum geometric tensor (QGT)~\cite{cheng2010quantum} using gauge-invariant projectors, and provide useful formulas for computation.
Quantum geometry describes the change of occupied wave functions $\uk$ in the momentum space.
The central quantity is the $d \times d$ Abelian quantum geometric tensor $\qgtmat$, whose entries are given by
\ba
\qgt = Tr[\dm \ukdag (\id - \proj) \dn \uk],
\quad
\dm \uk \equiv \lim_{\vep \rightarrow 0} \frac{\mc{U}(\bk + \vep \ehat_{\mu}) - \uk}{\vep}.
\label{seq:qgt}
\ea
Note that the derivatives in this formula cannot be numerically implemented due to the gauge freedom of $\uk$.
We handle this problem by obtaining an expression in terms of the gauge-invariant projectors as follows:
\ba
\qgt
&= Tr[\dm \ukdag (\id - \proj) \dn \uk \ukdag \uk] \quad (\because \ukdag \uk = \id)
\nn
&= Tr[\uk \dm \ukdag (\id - \proj) \dn \uk \ukdag] \quad (\because Tr[AB]=Tr[BA])
\nn
&= Tr[(\dm \proj - \dm \uk \ukdag) (\id - \proj) (\dn \proj - \uk \dn \ukdag)]
\nn
&= Tr[\dm \proj (\id - \proj) \dn \proj],
\ea
where the last line follows from
\ba
\proj \uk = \uk, \quad \ukdag \proj = \ukdag.
\ea
We can further simplify the this expression by noting
\ba
\proj^2 = \proj
\Rightarrow
\dm \proj \proj + \proj \dm \proj = \dm \proj
\Rightarrow
\dm \proj (\id - \proj) = \proj \dm \proj.
\label{seq:dproj}
\ea
Therefore, we obtain the computable expression
\ba
\qgt = Tr[\proj \dm \proj \dn \proj].
\label{seq:qgtproj}
\ea

The geometric interpretation of $\qgtmat$ follows from its relation to the quantum metric $\gmat$~\cite{provost1980riemannian,resta2011insulating} and the Berry curvature $\fmat$~\cite{berry1984quantal}.
To understand the quantum metric, let us introduce the Hilbert-Schmidt quantum distance
\ba
\qdist = \nocc - Tr[\proj P(\bk')],
\ea
which is a natural geometric measure of the dissimilarity between the occupied eigenstates $\uk$ and $\mc{U}(\bk')$.
Let us introduce some properties of the quantum distance.
Firstly, $0 \leq \qdist \leq \nocc$, which follows from
\ba
\qdist
= \nocc - Tr[\ukdag P(\bk') \uk]
= \nocc - \sum_n \bra{u_n(\bk)} P(\bk') \ket{u_n(\bk)}.
\ea
Since the eigenvalues of $P(\bk')$ are either $0$ or $1$, $0 \leq \bra{u_n(\bk)} P(\bk') \ket{u_n(\bk)} \leq 1$, we obtain the desired inequality.
Secondly, $\qdist = 0~(\nocc)$ if and only if the projectors are identical (orthogonal).
To prove this statement, let us introduce two useful theorems.
\textit{Theorem 1.---}
Let us consider a Hermitian matrix $A \psd 0$, where $\pdef (\psd)$ denotes positive (semi)definiteness.
Then,
\ba
\bra{v} A \ket{v} = 0
\lrar
A \ket{v} = 0
\lrar
\bra{v} A = 0
\label{seq:psdnull}
\ea

%
\textit{Proof for Theorem 1.---}
Let us denote the eigensystem of $A$ as $A\ket{a} = a\ket{a}$, where $a \geq 0$.
Then, the following matrix $B = \sum_a \sqrt{a} \kbr{a}{a}$ is Hermitian, and satisfies $A=B^{\dg} B$.
Thus, $\bra{v}A\ket{v}=0
\Leftrightarrow |B \ket{v}|^2 = 0
\Leftrightarrow  B\ket{v} = 0
\Rightarrow A \ket{v} = 0$.
Moreover, $A\ket{v} = 0 \Rightarrow \bra{v} A \ket{v} = 0$.
Taking a Hermitian conjugate completes the proof. $\square$

\textit{Theorem 2.---}
Let $P = \sum_p \lambda (p) \kbr{p}{p}$ and $Q = \sum_q \tau (q) \kbr{q}{q}$ denote positive semidefinite Hermitian matrices.
%
\ba
Tr[P Q] = 0 \lrar PQ=QP=0.
\label{seq:orthproj}
\ea

%
\textit{Proof for Theorem 2.---}
From Theorem 1, we find
\ba
Tr[PQ] 
&= 0
\nn
& \lrar \tau(q) \bra{q} P \ket{q} = 0~(^{\forall} q)
\nn
& \lrar
\tau(q) P \ket{q} = 0,
\quad
\tau(q) \bra{q} P = 0~(^{\forall} q)
\nn
& \rar
\sum_q P \tau(q) \kbr{q}{q} = \sum_q \tau(q) \kbr{q}{q} P = 0
\nn
& \lrar
PQ = QP = 0.
\ea
Moreover, $PQ = QP = 0 \rar Tr[PQ] = 0$ is trivial. $\square$

Using these results, we find that the condition for maximal quantum distance is given by
\ba
\qdist = \nocc
\lrar Tr[\proj P(\bk')] = 0
\lrar \proj P(\bk') = P(\bk') \proj = 0,
\ea
On the other hand, the condition for vanishing quantum distance is given by
\ba
\qdist = 0
\lrar Tr[(\id -\proj) P(\bk')] = 0
\lrar P(\bk') = \proj P(\bk') = P(\bk') \proj
\rar P(\bk') = \proj,
\ea
where we used the fact that $\id - \proj$ is also a projector in the first step; in the final step, we interchanged the $\bk$ and $\bk'$ indices.
Since $\proj = P(\bk')$ trivially implies $\qdist = 0$, we find that
\ba
\qdist = 0 \lrar \proj = P(\bk').
\ea
%

%
The quantum metric characterizes the local geometry imposed by the Hilbert-Schmidt quantum distance as $ds^2 = \qm dk_{\mu} dk_{\nu}$.
Explicitly, 
\ba
\qm
= \hf \frac{\der^2 s(\bk, \bk')^2}{\der k'_{\mu} \der k'_{\nu}} \Big|_{\bk' = \bk}
= -\hf Tr[\proj \dm \dn \proj].
\ea
To simplify this expression, note that
\ba
& Tr[\proj^2] = Tr[\proj] = \nocc
\nn
& \rar
Tr[\dm \dn \proj \proj + \dm \proj \dn \proj + \dn \proj \dm \proj + \proj \dm \dn \proj] = 0
\nn
& \lrar Tr[\proj \dm \dn \proj] = -Tr[\dm \proj \dn \proj] \quad (\because Tr[AB] = Tr[BA]).
\ea
Thus, the quantum metric is given by
\ba
\qm = \hf Tr[\dm \proj \dn \proj].
\ea
Crucially, the quantum metric is the real part of the quantum geometric tensor.
To prove this, first note that $\qgtmat = \qgtmat^{\dg}$, which follows from \eq{seq:qgt}:
\ba
\qgt^*
= Tr[\dm \uk^{T} (\id - \proj^T) \dn \uk^{*}]
= Tr[\dn \ukdag (\id - \proj) \dm \uk]
= \mf{G}_{\nu\mu}(\bk),
\ea
where we used $Tr[A]=Tr[A^T]$ to obtain the third equality.
Thus, we obtain
\ba
Re[\qgt]
&= \frac{\qgt + \mf{G}_{\nu\mu}(\bk)}{2} \quad (\because \qgtmat = \qgtmat^{\dg})
\nn
&= \hf Tr[\proj (\dm \proj \dn \proj + \dn \proj \dm \proj)]
\nn
&= \hf Tr[\dm \proj (\id - \proj) \dn \proj + \dm \proj \proj \dn \proj] \quad (\because Tr[AB] = Tr[BA])
\nn
&= \qm.
\ea

The quantum metric, in some sense, contains the information of the amplitude of the eigenvectors $\uk$.
On the other hand, the Berry curvature $\fmat$ characterizes the geometric phase induced by the parallel transport defined by the Abelian Berry connection
\ba
\mc{A}_{\mu} (\bk) = i Tr[\ukdag \dm \uk].
\ea
Explicitly,
\ba
\bcurv
&= \dm \mc{A}_{\nu} (\bk) - \dn \mc{A}_{\mu} (\bk)
\nn
&= i Tr[\dm \ukdag \dn \uk + \ukdag \dm \dn \uk- \dn \ukdag \dm \uk- \ukdag \dm \dn \uk]
\nn
&= i Tr[\dm \ukdag \dn \uk - \dn \ukdag \dm \uk].
\ea
Interestingly, the imaginary part of the quantum geometric tensor is related to the Berry curvature by
\ba
Im[\qgt]
&= \frac{\qgt - \mf{G}_{\nu\mu}(\bk)}{2i} \quad (\because \qgtmat = \qgtmat^{\dg})
\nn
&= \frac{Tr[\dm \ukdag(\id - \proj) \dn \uk - \dn \ukdag (\id - \proj) \dm \uk]}{2i}
\nn
&= -\hf \bcurv + \frac{i}{2} Tr[\dm \ukdag \uk \ukdag \dn \uk - \dn \ukdag \proj \dm \uk]
\nn
&= -\hf \bcurv + \frac{i}{2} Tr[\ukdag \dm \uk \dn \ukdag \uk - \dn \ukdag \proj \dm \uk] 
\nn
&= -\hf \bcurv \quad (\because Tr[AB] = Tr[BA]).
\ea
Note that the fourth equality follows from
\ba
\ukdag \uk = \id
\rar
\dm \ukdag \uk + \ukdag \dm \uk = 0
\lrar
\dm \ukdag \uk = -\ukdag \dm \uk. 
\ea
Thus, the following relations succinctly show the geometric aspects of $\qgtmat$
\ba
\qgtmat = \gmat - \frac{i}{2} \fmat,
\quad \gmat = Re[\qgtmat],
\quad \fmat = -2 Im[\qgtmat],
\label{seq:qgtrel}
\ea
which can be computed by numerically differentiating the projectors as~\cite{herzog2022superfluid}
\ba
\qgt = Tr[\proj \dm \proj \dn \proj],
\quad \qm = \hf Tr[\dm \proj \dn \proj],
\quad \bcurv = i Tr \big[\proj [\dm \proj, \dn \proj] \big],
\ea
where the expression for the Berry curvature follows from \eq{seq:qgtproj} and \eq{seq:qgtrel}.
%

%

\tocless{\subsubsection{Band-resolved quantum geometric tensor}
	\label{app:fid}}{}
Another important quantity which is closely related to quantum geometry is the \textit{band-resolved} QGT~\cite{watanabe2021chiral,kitamura2022quantum}, also known as the \textit{fidelity tensor}~\cite{jozsa1994fidelity,hwang2021geometric}, which characterizes the local interband transition probability through
\ba
p(m, \bk; n, \bk') = |\brkr{u_n(\bk')}{u_m(\bk)}|^2,
\quad
dp = \fid dk_{\mu} dk_{\nu},
\ea
where $m \neq n$.
Note that the linear term of $dp$ vanishes, due to $\brkr{u_n(\bk)}{u_m(\bk)} = 0$.
Following
\ba
dp
&=
\hf \frac{\der^2 p(m, \bk; n, \bk')}{\der k'_{\mu} \der k'_{\nu}} \Big|_{\bk' = \bk} dk_{\mu} dk_{\nu}
\nn
&= \hf \big( \brkr{\dm u_n(\bk)}{u_m(\bk)} \brkr{u_m(\bk)}{\dn u_n(\bk)} + (\mu \leftrightarrow \nu)\big) dk_{\mu} dk_{\nu}
\nn
&= \brkr{\dm u_n(\bk)}{u_m(\bk)} \brkr{u_m(\bk)}{\dn u_n(\bk)} dk_{\mu} dk_{\nu},
\ea
one obtains the band-resolved QGT
\ba
\fid = \brkr{\dm u_n(\bk)}{u_m(\bk)} \brkr{u_m(\bk)}{\dn u_n(\bk)},
\ea
which is a product of the non-Abelian interband Berry connections.
Using the projectors of the bands $P_{n}(\bk) = \kbr{u_n(\bk)}{u_n(\bk)}$, one obtains the following projector formula
\ba
\fid
&= Tr \big[\kbr{u_n(\bk)}{\dm u_n(\bk)}P_m(\bk) \kbr{\dn u_n(\bk)}{u_n(\bk)} \big]
\nn
&= Tr \big[(\dm P_n(\bk) - \kbr{\dm u_n(\bk)}{u_n(\bk)}) P_m(\bk) (\dn P_n(\bk) - \kbr{u_n(\bk)}{\dn u_n(\bk)}) \big]
\nn
&= Tr[\dm P_n(\bk) P_m(\bk) \dn P_n(\bk)]. \quad (\because \bra{u_n(\bk)} P_m(\bk) = 0, P_m(\bk) \ket{u_n(\bk)} = 0)
\ea
Note that the name \textit{band-resolved QGT} originates from
\ba
\qgt = \sum_{m>\nocc} \sum_{n \leq \nocc} \fid.
\ea

\section{The spin excitation Hamiltonian}
\label{app:Heff}
In this Appendix, we derive the spin excitation Hamiltonian in detail.
We denote the saturated ferromagnetic state as
\ba
\gs = \prod_{n}^{\nocc} \prod_{\bk}^{\nc} \pdag_{n \bk \ua} \vac,
\quad
c_{\bk \al \sg} \vac = 0,
\ea
and define spin excitations with momentum $\bQ$ as 
\ba
\ket{\bQ} =
\sum_{\bk}^{\nc} \sum_{\al}^{\norb} \sum_n^{\nocc}
z_{\bk \al n}(\bQ) \cdag_{\bk+\bQ \al \da} \psi_{n \bk \ua} \gs
=
\Pdag \gs,
\label{seq:Pdag}
\ea
which is a weighted sum of scatterings from the $n$-th occupied $\ua$-spin band to all the unoccupied orbitals $\al$.
The stable excitations are obtained by properly fixing $z_{\bk \al n}(\bQ)$.
Although the most physical definition is
\ba
[H,\Pdag] \gs = \Emag \ket{\bQ},
\label{seq:desired}
\ea
this equation does not have an exact solution in general~\cite{solyom2010fundamentals} unless the system is at half-filling.
As a resolution, we define the spin excitation energy as
\ba
\Emag = \frac{\braq H \ketq} {\brkr{\bQ}{\bQ}}
-
\brao H \keto,
\label{seq:Emag}
\ea
and perform a variation with respect to $z^*_{\bk \al n}(\bQ)$~\cite{shankar2012principles,griffiths2018introduction,wu2020quantum}.
To calculate the many-body correlations, we us introduce Wick's theorem of normal ordering \textit{with respect to} $\gs$~\cite{wick1950evaluation,giuliani2008quantum}.
Let us focus on the single particle operators $\pdag_{n\bk\ua}$, where we crucially restrict the band indices to $n = 1, \dots, \nocc$.
We define the normal order with respect to $\gs$ by permuting operators that annihilate $\gs$ to the right with a fermionic sign.
For a product of two operators, for instance, we have
\ba
&
:\pdag_{n_1 \bk_1 \ua} \psi_{n_2 \bk_2 \ua}:
=
-\psi_{n_2 \bk_2 \ua} \pdag_{n_1 \bk_1 \ua},
\quad
:\psi_{n_1 \bk_1 \ua} \pdag_{n_2 \bk_2 \ua}:
=
\psi_{n_1 \bk_1 \ua} \pdag_{n_2 \bk_2 \ua},
\nn
& :\psi_{n_1 \bk_1 \ua} \psi_{n_2 \bk_2 \ua}:
=
\psi_{n_1 \bk_1 \ua} \psi_{n_2 \bk_2 \ua},
\quad
:\pdag_{n_1 \bk_1 \ua} \pdag_{n_2 \bk_2 \ua}:
=
\pdag_{n_1 \bk_1 \ua} \pdag_{n_2 \bk_2 \ua}.
\ea
Importantly, the expectation value of normal ordered operators on $\gs$ vanish.
Let us define the contraction between two operators as
\ba
C(A B) = AB - :AB: = \expv{AB},
\ea
where $A$ and $B$ are either $\psi_{n \bk \ua}$ or $\pdag_{n \bk \ua}$, and $\expv{AB} = \bra{\Phi} AB \gs$.
Thus, the contraction is either $0$ or $1$.
Wick's theorem of normal ordering is the following operator identity
\ba
A_1 A_2 \dots A_N = :A_1 A_2 \dots A_N:
+
: \text{all possible contractions} :,
\ea
that is, all the terms containing $1, 2, \dots$ contractions.
Note that one must perform a signful permutation to bring two operators next to each other before performing the contraction.
Let us not compute \eq{seq:Emag}.
We begin with
\ba
I
&= \bra{\bQ} H \ket{\bQ}
\nn
&= \bra{\bQ} H_{\ua} + H_{\da} + H_{int} \ket{\bQ}
\nn
&= I_{\ua} + I_{\da} + I_{int}.
\ea
For later convenience, we express the Hamiltonians above as
\ba
&
H_\sg = \sum_{\bk \al \be} h(\bk)_{\al \be} \cdag_{\bk \al \sg} c_{\bk \be \sg} = \sum_{n \bk} E_n(\bk) \pdag_{n \bk \sg} \psi_{n \bk \sg},
\nn
& H_{int}
= \uc \sum_{\bk \bk' \bq \al} \cdag_{\bk + \bq \al \ua} \cdag_{\bk' - \bq \al \da} c_{\bk' \al \da} c_{\bk \al \ua}
= \uc \sum_{\bk \bk' \bq \al} \sum_{l_1 l_2} \ls{\al} \brk{u_{l_1}(\bk + \bq)}{u_{l_2}(\bk)}_{\al}  \pdag_{l_1 \bk+\bq \ua} \cdag_{\bk' - \bq \al \da} c_{\bk' \al \da} \psi_{l_2 \bk \ua}.
\ea
Proceeding with $I_\ua$, we obtain
\ba
I_\ua
&= \sum_{\bk_1 \al_1 n_1} \sum_{\bk_2 \al_2 n_2}
z^*_{\bk_1 \al_1 n_1}(\bQ) z_{\bk_2 \al_2 n_2}(\bQ) \sum_{n \bk} E_n(\bk)
\expv{
	\pdag_{n_1 \bk_1 \ua} c_{\bk_1 + \bQ \al_1 \da}
	\pdag_{n \bk \ua} \psi_{n \bk \ua}
	\cdag_{\bk_2 + \bQ \al_2 \da} \psi_{n_2 \bk_2 \ua}
}
\nn
&= \sum_{\bk_1 \al_1 n_1 n_2} 
z^*_{\bk_1 \al_1 n_1}(\bQ) z_{\bk_1 \al_1 n_2}(\bQ) \sum_{n \bk} E_n(\bk)
\expv{
	\pdag_{n_1 \bk_1 \ua} \pdag_{n \bk \ua} \psi_{n \bk \ua} \psi_{n_2 \bk_1 \ua}
}.
\ea
Considering only finite contributions, we can restrict the band index as $n = 1, \dots, \nocc$, which allows us to use Wick's theorem.
Thus, we find
\ba
I_\ua = \sum_{\bk \al n}
|z_{\bk \al n}(\bQ)|^2 \sum_{l \bq} E_l(\bq)
-\sum_{\bk \al n} |z_{\bk \al n}(\bQ)|^2
E_{n}(\bk).
\ea
Calculating the second term, we obtain
\ba
I_\da
&= \sum_{\bk_1 \al_1 n_1} \sum_{\bk_2 \al_2 n_2}
z^*_{\bk_1 \al_1 n_1}(\bQ) z_{\bk_2 \al_2 n_2}(\bQ)
\sum_{\bk \al \be} h(\bk)_{\al \be}
\expv{
	\pdag_{n_1 \bk_1 \ua} c_{\bk_1 + \bQ \al_1 \da}
	\cdag_{\bk \al \da} c_{\bk \be \da}
	\cdag_{\bk_2 + \bQ \al_2 \da} \psi_{n_2 \bk_2 \ua}
}
\nn
&= \sum_{\bk_1 n_1 \al_1} \sum_{\bk_2 n_2 \al_2} z^*_{\bk_1 \al_1 n_1}(\bQ) z_{\bk_2 \al_2 n_2}(\bQ) \sum_\al h(\bk_2 + \bQ)_{\al \al_2}
\expv{
	\pdag_{n_1 \bk_1 \ua} c_{\bk_1 + \bQ \al_1 \da} \cdag_{\bk_2 + \bQ \al \da} \psi_{n_2 \bk_2 \ua}
}
\nn
&= \sum_{\bk \al \be n} z^*_{\bk \al n}(\bQ) z_{\bk \be n}(\bQ) h(\bk + \bQ)_{\al \be}.
\ea
The third term, $I_{int}$, is expressed as
\ba
I_{int}
&=
\uc \sum_{\bk_1 \al_1 n_1} \sum_{\bk_2 \al_2 n_2}
z^*_{\bk_1 \al_1 n_1}(\bQ) z_{\bk_2 \al_2 n_2}(\bQ)
\sum_{\bk \bk' \bq \al} \sum_{l_1 l_2} \ls{\al} \brk{u_{l_1}(\bk + \bq)}{u_{l_2}(\bk)}_{\al}
\nn
& \quad \times
\expv{
	\pdag_{n_1 \bk_1 \ua} c_{\bk_1 + \bQ \al_1 \da}
	\pdag_{l_1 \bk+\bq \ua} \cdag_{\bk' - \bq \al \da} c_{\bk' \al \da} \psi_{l_2 \bk \ua}
	\cdag_{\bk_2 + \bQ \al_2 \da} \psi_{n_2 \bk_2 \ua}
}
\nn
&=
-\uc \sum_{\bk_1 \al_1 n_1} \sum_{\bk_2 \al_2 n_2}
z^*_{\bk_1 \al_1 n_1}(\bQ) z_{\bk_2 \al_2 n_2}(\bQ)
\sum_{\bk \bq l_1 l_2}
\ls{\al_2} \brk{u_{l_1}(\bk + \bq)}{u_{l_2}(\bk)}_{\al_2}
\nn
& \quad \times
\expv{
	\pdag_{n_1 \bk_1 \ua} c_{\bk_1 + \bQ \al_1 \da}
	\pdag_{l_1 \bk+\bq \ua} \cdag_{\bk_2 + \bQ - \bq \al_2 \da} \psi_{l_2 \bk \ua} \psi_{n_2 \bk_2 \ua}
}
\nn
&=
\uc \sum_{\bk_1 \bk_2 \al_1 n_1 n_2} z^*_{\bk_1 \al_1 n_1}(\bQ) z_{\bk_2 \al_1 n_2} (\bQ)
\sum_{\bk l_1 l_2} \ls{\al_1} \brk{u_{l_1}(\bk + \bk_2 - \bk_1)}{u_{l_2}(\bk)}_{\al_1}
\expv{
	\pdag_{n_1 \bk_1 \ua} \pdag_{l_1 \bk + \bk_2 - \bk_1 \ua} \psi_{l_2 \bk \ua} \psi_{n_2 \bk_2 \ua}
}.
\ea
Again, we may restrict the $l_1, l_2$ indices to $1, \dots, \nocc$.
Applying Wick's theorem, we obtain
\ba
I_{int}
= \uc \sum_{\bk \al n}
\Big(
|z_{\bk \al n}(\bQ)|^2 \sum_{\bq l} \ls{\al} \brk{u_l(\bq)}{u_l(\bq)}_{\al}
- \sum_{\bk' n'} z^*_{\bk \al n}(\bQ) z_{\bk' \al n'}(\bQ)
\ls{\al} \brk{u_{n'}(\bk')}{u_n(\bk)}_\al
\Big).
\ea
The two remaining terms are given by
\ba
& \expv{H} = \sum_n^{\nocc}\sum_\bk E_n(\bk),
\nn
& \brk{\bQ}{\bQ} = 
\sum_{\bk_1 \al_1 n_1} \sum_{\bk_2 \al_2 n_2}
z^*_{\bk_1 \al_1 n_1}(\bQ) z_{\bk_2 \al_2 n_2}(\bQ)
\expv{
	\pdag_{n_1 \bk_1 \ua} c_{\bk_1 + \bQ \al_1 \da}
	\cdag_{\bk_2 + \bQ \al_2 \da} \psi_{n_2 \bk_2 \ua}
}
= \sum_{\bk \al n} |z_{\bk \al n}(\bQ)|^2.
\ea
Using these results, we obtain for \eq{seq:Emag}
\ba
\Emag \sum_{n \al \bk}|z_{\bk \al n}(\bQ)|^2
&=
\sum_{n \bk \al \be} z^*_{\bk \al n}(\bQ) h(\bk + \bQ)_{\al \be} z_{\bk \be n}(\bQ)
-
\sum_{n \al \bk} |z_{\bk \al n}(\bQ)|^2 E_n(\bk)
\nn
&+
\Uc \sum_{\al} \sum_{n \bk}|z_{\bk \al n}(\bQ)|^2 \sum_{l \bq} \ls{\al} \brkr{u_{l}(\bq)}{u_{l}(\bq)}_{\al}
-
\Uc \sum_{n n' \al \bk \bk'} z^*_{\bk \al n}(\bQ) z_{\bk' \al n'}(\bQ) \ls{\al} \brkr{u_{n'}(\bk')}{u_n(\bk)}_{\al}
.
\ea
Next, we take a partial derivative with respect to $z^*_{\bk \al n}(\bQ)$ and obtain
\ba
\Emag z_{\bk \al n}(\bQ)
&=
\sum_{\be} h(\bk + \bQ)_{\al \be} z_{\bk \be n}(\bQ)
-
z_{\bk \al n}(\bQ) E_n(\bk)
+
\Uc z_{\bk \al n}(\bQ) \sum_{l \bq} \ls{\al} \brkr{u_l(\bq)}{u_l(\bq)}_{\al}
\nn
&-
\Uc \sum_{n' \bk'} z_{\bk' \al n'}(\bQ) \ls{\al} \brkr{u_{n'}(\bk')}{u_n(\bk)}_{\al}.
\ea
This equation can be solved by converting it to an eigenvalue problem~\cite{solyom2010fundamentals},
\ba
\Emag z_{\bk \al n}(\bQ) = \sum_{\bk' \be n'} \Heff z_{\bk' \be n'}(\bQ),
\label{seq:Heig}
\ea
where
\ba
\Heff
&=
h(\bk+\bQ)_{\al \be} \delta_{\bk \bk'} \delta_{n n'}
-
E_n(\bk) \delta_{\bk \bk'} \delta_{\al \be} \delta_{n n'}
+
\Uc \sum_{\bq l} \ls{\al} \brkr{u_l(\bq)}{u_l(\bq)}_{\al} \delta_{\bk \bk'} \delta_{\al \be} \delta_{n n'}
\nn
&-
\Uc \ls{\al} \brkr{u_{n'}(\bk')}{u_n(\bk)}_{\al} \delta_{\al \be}
\label{seq:Heff}
\ea
is the spin excitation Hamiltonian whose rows and columns are indexed by $\bk \al n$ and $\bk' \be n'$, respectively.
Note that the only approximation we used is the definition of the spin excitation energy in \eq{seq:Emag}, which is inevitable due to the failure of \eq{seq:desired}.
Other than this, the spin excitation energy is exact, which we explicitly proved using Wick's theorem.

\section{Spin excitations and the Goldstone mode}
\label{app:magprop}
%
In this Appendix, we introduce noteworthy aspects of the spin excitation spectrum.
As illustrated in \fig{fig1}(c) of the main text, spin excitations consist of two parts.
The first part is the quasiparticle excitations which form the Stoner continuum, whose energy scales linearly with $U$.
The second part encompasses the $\norb$ collective excitations, also known as \textit{spin waves}, which exist below the Stoner continuum.
Contrary to the quasiparticle excitations, the spin wave energy reaches an upper bound determined by $h(\bk)$ as $U$ increases.
The spin waves further consist of $\norb-1$ gapped modes and $1$ gapless mode, where the latter corresponds to a Goldstone boson generated by spontaneous symmetry breaking of the $SU(2)$ group.
Let us solve for the gapless mode at $\bQ = \bb 0$.
First, note that $\mc{E}_G (\bb 0) = 0$ indicates that the Goldstone mode at $\bQ = \bb 0$ corresponds to another ground state of $H$ with $S_{tot}^z = \smax - 1$.
Since the ground states are given by $S^-_{tot}\ket{GS}$, we can intuitively predict that $\ket{\bQ}$ will be given by $S_{tot}^{-} \gs$.
Let $\psi_{n \bk \sg}^\dg = \sum_{\al}^{\norb} \ket{u_n(\bk)}_{\al} \cdag_{\bk \al \sg}$ denote the creation operator of the $n$-th band of the kinetic Hamiltonian.
In terms of this operator, we obtain
\ba
S_{tot}^-
&=
\sum_{i \al} \cdag_{i \al \da} c_{i \al \ua}
=
\sum_{\bk \al} \cdag_{\bk \al \da} c_{\bk \al \ua}
=
\sum_{\bk \al m n} \psi^{\dg}_{m \bk \da} \psi_{n \bk \ua} \cdot \ls{\al} \brk{u_m (\bk)}{u_n (\bk)}_{\al}
=
\sum_{\bk n} \psi_{n \bk \da}^{\dg} \psi_{n \bk \ua},
\label{seq:sminus}
\ea
where the sum over $\al, m, n$ runs through $1, \dots, \norb$.
Combined with \eq{seq:Pdag}, we obtain
\ba
z_{\bk \al n} (\bb 0) = \ket{u_n (\bk)}_\al
\label{seq:zg}
\ea
for the gapless mode.
We verify that this does indeed correspond to the gapless mode by substituting into \eq{seq:Heig} as follows.
\ba
\sum_{n' \bk' \be} \mc{H}_{\bk \al n, \bk' \be n'}^{SE} (\bb 0) \ket{u_{n'} (\bk ')}_{\be}
&=
\sum_{n' \bk' \be}
\Bigl[
\bigl\{h(\bk)_{\al \be} - E_n (\bk) \delta_{\al \be}
\bigr\}
\delta_{\bk \bk'} \delta_{n n'}
+
\uc
\bigl\{
\sum_{\bq l} \ls{\al} \brkr{u_l (\bq)}{u_l (\bq)}_{\al} \delta_{\bk \bk'} \delta_{n n'}
\nn
&-
\ls{\al} \brkr{u_{n'} (\bk')}{u_n (\bk)}_{\al}
\bigr\}
\delta_{\al \be}
\Bigr]
\ket{u_{n'} (\bk')}_{\be}
\nn
&=
\sum_\be h(\bk)_{\al \be} \ket{u_n(\bk)}_{\be} - E_n (\bk) \ket{u_n (\bk)}_{\al}
+
\uc \sum_{\bq l} \ls{\al} \brkr{u_l (\bq)}{u_l (\bq)}_{\al}\ket{u_n (\bk)}_{\al}
\nn
&-
\uc \sum_{n' \bk'} \ls{\al} \brkr{u_{n'} (\bk')}{u_n (\bk)}_{\al} \ket{u_{n'}(\bk')}_{\al}
\nn
&=
E_n(\bk) \ket{u_n (\bk)}_{\al} - E_n (\bk) \ket{u_n (\bk)}_{\al}
\nn
&+
\uc
\Big(
\sum_{\bq l} \ls{\al} \brkr{u_l (\bq)}{u_l (\bq)}_{\al}
-
\sum_{n' \bk'} \ls{\al} \brk{u_{n'} (\bk')}{u_{n'}(\bk')}_{\al}
\Big)
\ket{u_n(\bk)}_{\al}
\nn
&= 0,
\label{seq:zmagCal}
\ea
where $n', l$ are summed over $1, \dots, \nocc$.
Thus, \eq{seq:zg} solves the Goldstone mode mode at $\bQ = \bb 0$.
Note that spin excitations with negative energy imply that we have assumed the wrong ground state when calculating $\Hse$.
Conversely, for a saturated ferromagnet, the gapless magnon mode is the nondegenerate ground state of $\Hse(\bb 0)$, since the other spin excitations are gapped.
Furthermore, $\Eg > 0$ is required, since any ground state of $H$ must have maximal total spin, which is impossible for an arbitrary $\bQ$.
This implies that $\bQ = \bb 0$ corresponds to the global minimum of $\Em$.
Expanding around $\Gamma$, we obtain
\ba
\Em 
&= \mc{E}_G(\bb 0) + Q_\mu \der_\mu \Em|_{\bQ = \bb 0} + \frac{Q_\mu Q_\nu}{2} \der_\mu \der_\nu \Em |_{\bQ = \bb 0}
+
O(Q^3),
\label{seq:eGamma}
\ea
where the repeated indices are summed over.
Thus, the linear term must vanish, and the spin stiffness $\Dmag \equiv \der_\mu \der_\nu \Em |_{\bQ = \bb 0}$ must be a positive definite tensor.
Using the Feynman-Hellmann theorem~\cite{guttinger1932verhalten,feynman1939forces}, we show that the first condition is always satisfied:
\ba
\der_\mu \Em|_{\bQ = \bb 0}
&=
\frac{\sum_{\bk \al n}\sum_{\bk' \be n'} \der_\mu \Hse (\bQ) |_{\bQ = \bb 0} z^*_{\bk \al n}(\bb 0) z_{\bk' \be n'}(\bb 0)}{\sum_{n \al \bk} |z_{n \al \bk}(\bb 0)|^2}
\nn
&=
\den{\ntot}
{\sum_{n \bk} \bra{u_n (\bk)}
	\frac{\der h(\bk + \bQ)}{\der Q_\mu} \bigg|_{\bQ = \bb 0}
	\ket{u_n (\bk)}}
\nn
&=
\den{\ntot} \sum_{n \bk} \bra{u_n(\bk)} \der_\mu h(\bk) \ket{u_n (\bk)}
\nn
&=
\den{\ntot} \sum_{n \bk} \der_\mu E_n(\bk)
\nn
&= 0,
\label{seq:linear}
\ea
where the first and fourth equality comes from the Feynman-Hellman theorem, and the last equality comes from the fact that $E_n (\bk)$ is periodic throughout the 1st Brillouin zone (BZ).
Therefore, a necessary but not sufficient condition for saturated ferromagnetism is that $\Dmag$ is a positive definite tensor.
%

\section{Approximation to the Goldstone mode}
\label{app:approx}
In this Appendix, we derive the perturbative expansion of the Goldstone mode of the spin excitation spectrum, and discuss its properties.
Let us describe the eigensystem of the spin excitation Hamiltonian as
\ba
\Hse(\bQ) \ket{z_n(\bQ)} = \mc{E}_n(\bQ) \ket{z_n(\bQ)} \quad (n = 1, \dots, \ntot \norb).
\ea
Due to the complexity of $\Hse(\bQ)$, it is impossible to find an analytic solution.
Thus, we obtain a perturbative expression by separating the terms as
\ba
\Hse(\bQ) = \Hse(\bb 0) + V(\bQ),
\ea
where the unperturbed Hamiltonian is given by
\ba
\Hse_{\bk \al n, \bk' \be n'}(\bb 0) = [h(\bk)_{\al \be}
-
E_n(\bk)\delta_{\al \be}] \delta_{\bk \bk'} \delta_{n n'}
+
\Uc \Big[ \sum_{\bq}^{\nc} \sum_l^{\nocc} \ls{\al} \brkr{u_l(\bq)}{u_l(\bq)}_{\al} \delta_{\bk \bk'} \delta_{n n'}
-
\ls{\al} \brkr{u_{n'}(\bk')}{u_n(\bk)}_{\al} \Big] \delta_{\al \be},
\ea
and the perturbation is given by
\ba
V_{\bk\al n, \bk' \be n'}(\bQ)
=
[h(\bk+\bQ)-h(\bk)]_{\al \be} \delta_{\bk \bk'} \delta_{n n'}.
\ea
Then, the Goldstone mode for a ferromagnetic state is given by
\ba
\Eg
&= \Egb^{(1)}(\bQ) + \Egb^{(2)}(\bQ) + \dots
\nn
&= \bra{z_1 (\bb 0)} V(\bb Q) \ket{z_1 (\bb 0)}
- \sum_{n>1} \frac{|\bra{z_1 (\bb 0)} V(\bQ) \ket{z_n (\bb 0)}|^2}{\mc{E}_n (\bb 0)} + \dots.
\label{seq:pert}
\ea
Note that such perturbative expansion is only justified when the unperturbed Hamiltonian is much larger than the perturbation.
Since the former contains $U$-linear terms, this analysis is valid in the $U \rightarrow \infty$ limit.
Even so, the first order term serves as a rigorous upper bound of $\Eg$ for any $U$.
To show this, note that
\ba
\Eg = \bra{z_1(\bQ)}\Hse(\bQ) \ket{z_1(\bQ)} \leq \bra{v} \Hse(\bQ) \ket{v}
\ea
for an arbitrary vector $\ket{v}$, since $\ket{z_1(\bQ)}$ is the true ground state eigenvector of $\Hse(\bQ)$.
Setting $\ket{v} = \ket{z_1(\bb 0)}$, we obtain the desired result.
%

\tocless{\subsection{First order perturbation theory}
	\label{appsub:first}}{}
In the strongly correlated limit, we have verified in various models that first order perturbation is sufficient for determining the stability of the true Goldstone model $\Eg$; in fact, in many cases, $\Eg$ saturates to $\Epert$ at small $\bQ$ when the \textit{minimal metric condition} is satisfied.
In more mathematical terms, resurgence does not occur from second order perturbation in this series.
Thus, we focus on the first order term.
Using $\ket{z_1(\bb 0)}_{\bk \al n} = \frac{\ket{u_n(\bk)}_{\al}}{\sqrt{\ntot}}$ for the (normalized) Goldstone mode, we obtain
\ba
\Epert
&= \bra{z_1(\bb 0)} V(\bQ) \ket{z_1(\bb 0)}
\nn
&=
\den{\ntot}
\sum_{\bk} \sum_n^{\nocc}
\bra{u_n (\bk)} h(\bk + \bQ)-h(\bk) \ket{u_n(\bk)}
\nn
&= \den{\ntot}\sum_{\bk} Tr[(h(\bk + \bQ) - h(\bk)) \proj].
\ea

%
We show that this term is always non-negative.
To do this, let us assume that $I_l \subset \{1, \dots, \norb\}~(l = 1, \dots, \ltot)$ indexes $|I_l|$ bands degenerate at every momentum, with energy $\Etd_l(\bk) = E_{\sum_{l' < l} |I_{l'}|+1}(\bk)$.
Moreover, we assume that $\nocc = \sum_{l=1}^{\lmax} |I_l|$.
For example, in a $3$- band system with $2$ occupied bands, which obeys $E_1(\bk) = E_2(\bk)$ at every $\bk$, then $I_1 = \{1,2\}$ with $\Etd_1(\bk) = E_1(\bk)$, and $I_2 = \{3\}$ with $\Etd_2(\bk) = E_3(\bk)$.
In addition, $\lmax = 1$ and $\ltot = 2$.
Then, the kinetic Hamiltonian is expressed as
\ba
h(\bk) = \sum_l \Etd_l (\bk) \Ptd_l (\bk) \quad (\Ptd_l(\bk) = \sum_{n \in I_l} \kbr{u_n(\bk)}{u_n(\bk)}).
\label{seq:hl}
\ea

We now claim that for any projector $\pcurv$ onto an $\nocc$-dimensional subspace,
\ba
Tr[h(\bk) \pcurv] \geq \sum_{n}^{\nocc} E_n(\bk),
\label{seq:minproj}
\ea
and that the equality saturates iff $\pcurv = \proj$.
The proof is as follows.
Let
\ba
\al_l \equiv Tr[\Ptd_l(\bk) \pcurv].
\ea
Since $\sum_l^{\ltot} \Ptd_l(\bk) = \id$, one has $\sum_{l}^{\ltot} \al_l = \nocc$.
Moreover, one finds
\ba
0 \leq \al_l = \sum_{n \in I_l} \bra{u_n(\bk)} \pcurv \ket{u_n(\bk)} \leq |I_l|.
\ea
Then, the left hand side of \eq{seq:minproj} simply reduces to
\ba
Tr[h(\bk) \pcurv]
= \sum_{l}^{\ltot} Tr[\Etd_l (\bk) \Ptd_l (\bk) \pcurv]
= \sum_{l}^{\ltot} \al_l \Etd_l(\bk).
\ea
Clearly, this is minimized iff $\al_l = |I_l|~(0)$ for $l \leq \lmax~(l > \lmax)$.
From \eq{seq:orthproj}, it readily follows that
\ba
\pcurv
= \pcurv \sum_{l}^{\ltot} \Ptd_l(\bk)
= \pcurv \sum_{l}^{\lmax} \Ptd_l(\bk)
= \pcurv \proj.
\ea
This implies that the quantum distance between two projectors with equal trace, $\pcurv$ and $\proj$, vanishes, which is equivalent to $\pcurv = \proj$.
$\square$
Applying this to $\Epert$, we obtain
\ba
\Epert
= \den{\ntot}\sum_{\bk} Tr[(h(\bk + \bQ) - h(\bk)) \proj]
\geq \den{\ntot}\sum_{n \bk} (E_n(\bk + \bQ) - E_n(\bk))
= 0,
\ea
where the lower bound saturates iff $P(\bk + \bQ) = \proj$ for every $\bk$.
In terns of quantum geometry, this means that $s(\bk, \bk + \bQ)^2 = 0$ for every $\bk$.
As discussed in the main text, the stability condition at $\bQ = \bb 0$ requires a positive definite quantum metric integral.
Expanding $\Epert$ up to quadratic terms, we obtain
\ba
\Epert
&=
\hf Q_\mu Q_\nu \dm \dn \Eg |_{\bQ = \bb 0} 
\nn
&=
\qmn \sumk \sumnn \bra{\un} \frac{\der^2 h(\bk + \bQ)}{\der Q_\mu \der Q_\nu} \bigg |_{\bQ = \bb 0} \ket{\un}
\nn
&=
\qmn \sumk \sumnn \bra{\un} \dm\dn h(\bk) \ket{\un}.
\label{seq:emdk}
\ea
Inserting a resolution of identity, we obtain
\ba
\Epert
&=
\qmn \sumk
\sumnn \summ \bra{\un} \Big[
\dm \dn \kbr{\um}{\um} \emm
\Big]
\ket{\un}
\nn
&=
\qmn \sumk \sumnn \summ
\bra{\un}
\nn
&
\Big[
\kbr{\dm\dn \um}{\um} \emm
+
\kbr{\dm \um}{\dn \um} \emm
+
\kbr{\dm \um}{\um} \dn \emm
\nn
&+
\kbr{\dn \um}{\dm \um} \emm
+
\kbr{\um}{\dm \dn \um}\emm
+
\kbr{\um}{\dm \um} \dn \emm
\nn
&+
\kbr{\dn \um}{\um} \dm \emm
+
\kbr{\um}{\dn \um} \dm \emm
+
\kbr{\um}{\um} \dm\dn \emm
\Big]
\ket{\un}
\nn
&=
\qmn \sumk \sumnn
\Big\{
\brk{\un}{\dm\dn \un} \enn
+
\brk{\un}{\dm \un} \dn \enn
+
\brk{\dm\dn \un}{\un} \enn
\nn
&+
\brk{\dm \un}{\un} \dn \enn
+
\brk{\un}{\dn \un} \dm \enn
+
\brk{\dn \un}{\un} \dm \enn
+
\dm \dn \enn
\nn
&+
\summ
\big(
\brk{\un}{\dm \um} \brk{\dn \um}{\un}
+
\brk{\un}{\dn \um}\brk{\dm \um}{\un}
\big) \emm
\Big\}.
\label{seq:empertII}
\ea
To simplify \eq{seq:empertII}, we use the following identities:
\ba
&
\brk{\um}{\un} = \delta_{\mu \nu},
\quad
\brk{\dm \um}{\un} + \brk{\um}{\dm \un} = 0, \quad \text{and}
\nn
&
\brk{\dm \dn \um}{\un} + \brk{\um}{\dm \dn \un} + \brk{\dm \um}{\dn \un} + \brk{\dn \um}{\dm \un} = 0.
\ea
Then, we obtain
\ba
\Epert
&=
\qmn \sumk \sumnn
\Big[
\dm \dn \enn
-
\big(
\brk{\dm \un}{\dn \un} + \brk{\dn \un}{\dm \un}
\big)
\enn
\nn
&+
\summ
\big(
\brk{\dm \um}{\un} \brk{\un}{\dn \um} + \brk{\dn \um}{\un} \brk{\un}{\dm \um}) \emm
\Big]
\nn
&=
\qmn \sumk \sumnn
\Big[
\dm\dn \enn
+
\summ
\big(
\brk{\dm \um}{\un} \brk{\un}{\dn\um} + \mnarrow
\big)
\emm
\nn
&-
\summ
\big(
\brk{\dm \un}{\um} \brk{\um}{\dn\un} + \brk{\dn \un}{\um} \brk{\um}{\dm \un}
\big)
\enn
\Big]
\nn
&=
\qmn \sumk \sumnn
\Big[
\dm\dn \enn
+
\summ
\big(
\brk{\dm \um}{\un} \brk{\un}{\dn\um} + \mnarrow
\big)
\emm
\nn
&-
\summ
\big(
\brk{\un}{\dm \um} \brk{\dn\um}{\un} + \brk{\un}{\dn\um} \brk{\dm\um}{\un}
\big)
\enn
\Big]
\nn
&=
\qmn \sumk
\Big[
\sumnn \dm\dn \enn
+
\summ \sumnn (\emm-\enn)
\big(
\brk{\dm \um}{\un} \brk{\un}{\dn\um} + \mnarrow
\big)
\Big],
\label{seq:EmpertIII}
\ea
where we have inserted a resolution of identity to obtain the second equality.
In terms of $\fid$, this reduces to
\ba
\Epert
&=
\qmn \sumk
\Big[
\sumnn \dm \dn \enn
+
\sum_m^{\norb}\sum_n^{\nocc} (\emm - \enn) (\fid + \chi^{mn}_{\nu \mu}(\bk))
\Big]
\nn
&=
\qmn \sumk
\Big[
\sumnn \dm \dn \enn
+
2 \sum_m^{\norb}\sum_n^{\nocc} (\emm - \enn) \fid
\Big],
\label{seq:fidFinal}
\ea
where we have interchanged the summation indices $\mu$ and $\nu$ of the last term to obtain the second equality.
We further note that $\fid=\chi_{\nu\mu}^{nm}(\bk)$.
Then, the terms with $m \in \{1, \dots, \nocc\}$ in the second summation become $0$ from the following relation:
\ba
&
\sum_{\mu \nu}^d \sum_{m}^{\nocc} \sum_n^{\nocc} (\emm-\enn) \fid Q_{\mu} Q_{\nu}
\nn
&=
\sum_{\mu \nu}^d \sum_{m}^{\nocc} \sum_n^{\nocc} (\enn-\emm) \chi_{\nu\mu}^{nm}(\bk) Q_{\mu} Q_{\nu} 
\nn
&=
-\sum_{\mu \nu}^d \sum_{m}^{\nocc} \sum_n^{\nocc} (\emm-\enn) \fid Q_{\mu} Q_{\nu}.
\ea
Thus, we finally arrive at
\ba
\Epert
=
\qmn \sumk
\Big[
\sumnn \dm \dn \enn
+
2 \sum_{m>\nocc}^{\norb}\sum_n^{\nocc} (\emm - \enn) \fid
\Big].
\ea

Note that the first term of this expression vanishes when $E_n(\bk)$ are smooth and periodic.
This is because integrating the derivative of a smooth periodic function over a period vanishes.
In such cases, it is clear that the fidelity tensor determines whether the spin stiffness can be positive definite.
On the other hand, when the occupied bands form a kink structure due to a band crossing with the unoccupied bands, the first term does not vanish.
This may lead one to think that the singularity of the band structure, regardless of quantum geometry, can stabilize ferromagnetism.
This is not the case, however, since a singularity in $E_n(\bk)$ necessarily causes a divergence in $\fid$.
We prove this by contradiction.
Without loss of generality, we set $\lmax = 1$ (see \eq{seq:hl}), and consider a band crossing at $\bk^*$.
Around this point, we assume that $\dm \Etd_1(\bk)$ is discontinuous, while $\Ptd_1(\bk)$ is smooth.
Let us now consider
\ba
h(\bk) \Ptd_1(\bk)
=
\Etd_1(\bk) \Ptd_1(\bk).
\ea
Since $h(\bk)$ is smooth, this expression must be smooth; moreover,
\ba
\dm(h(\bk) \Ptd_1(\bk))
=
\dm \Etd_1(\bk) \Ptd_1(\bk) + \Etd_1(\bk) \dm \Ptd_1(\bk)
\ea
must also be smooth.
However, $\dm \Etd_1(\bk) \Ptd_1(\bk)$ is discontinuous.
Thus, a kink in the band structure necessarily induces a singularity in the occupied band eigenstates.
Therefore, $\Epert > 0$ cannot occur without a nontrivial quantum geometry, even if the first term is finite.

\phantom{}

\tocless{\subsubsection{Upper bound of the spin stiffness}
	\label{appsub:dbound}}{}
In this Appendix, we prove that
\ba
\Eg|_{U < \infty} \leq \Eg |_{U \rightarrow \infty} \leq \Epert
\label{seq:mIneq}
\ea
implies
\ba
\ddmag (U < \infty) \leq \ddmag (U \rightarrow \infty) \leq \ddpert.
\label{seq:DIneq}
\ea
The proof follows directly in the following way.
Consider two smooth analytic functions $f$ and $g$, where $f(\bb 0) = g(\bb 0) = 0$ and $f(\bQ) \leq g(\bQ)$.
Let $F(\bQ) = g(\bQ) - f(\bQ) \geq 0$, with $F(\bb 0) = 0$.
This implies that Hessian of $F(\bQ)$ defined by $\mf{D}_{\mu \nu} \equiv \frac{\der^2 F(\bQ)}{\der Q_\mu \der Q_\nu}\big|_{\bQ = \bb 0}$ is positive semidefinite.
This implies that the diagonal entries are greater than or equal to $0$.
Applying this to \eq{seq:mIneq} finishes the proof.

\phantom{}

\tocless{\subsection{Second order perturbation theory}
	\label{appsub:second}}{}
In this Appendix, we discuss the condition where $\limu \Dmag = \Dpert$.
Starting from
\ba
\Eg
&= \Epert + \Esec + \dots
\nn
&= \bra{z_1 (\bb 0)} V(\bb Q) \ket{z_1 (\bb 0)}
- \sum_{n>1} \frac{|\bra{z_1 (\bb 0)} V(\bQ) \ket{z_n (\bb 0)}|^2}{\mc{E}_n (\bb 0)} + \dots,
\ea
we find that
\ba
D_{\mu\nu} = \Dpert +\Dsec,
\ea
where the higher order terms vanish due since they contain terms $\bra{z_n(\bb 0)}V(\bb 0) \ket{z_1(\bb 0)} = 0$.
Thus, we obtain
\ba
\limu \Dmag = \Dpert
\lrar
\limu \Dsec = 0,
\ea
since $\Dpert$ is independent of $U$.
To find this condition, note that
\ba
\limu \mc{E}_n(\bb 0)
\begin{cases}
	< \infty & (n \leq \norb) \\
	= \infty & (n > \norb)
\end{cases}
,
\ea
that is, the energy of Stoner continuum states diverge, while the magnon modes saturate to a finite energy.
This indicates that at $U \rightarrow \infty$, the $\norb$ magnon modes form a basis of the null space of the $U$-dependent part of $\Hse(\bb 0)$.
Note that this statement is only true in the strongly correlated limit, since $\ket{z_n(\bb 0)}~(n = 2, \dots, \norb)$ are a function of $U$ in general.
Thus, an equivalent condition for $\Dsec = 0$ may be obtained by expanding the second order perturbation up to quadratic order as
\ba
\limu \mc{E}_G^{(2)}(\bQ) 
&= -\hf Q_{\mu} Q_{\nu} \dm \dn \limu \sum_{n=2}^{\norb}  \frac{|\bra{z_1 (\bb 0)} V(\bQ) \ket{z_n (\bb 0)}|^2}{\mc{E}_n (\bb 0)} \Big|_{\bQ = \bb 0}
\nn
&= -\hf Q_{\mu} Q_{\nu} \limu \sum_{n=2}^{\norb} \frac{\bra{z_1(\bb 0)}\dm V(\bb 0) \ket{z_n(\bb 0)}\bra{z_n(\bb 0)}\dn V(\bb 0) \ket{z_1(\bb 0)} + (\mu \leftrightarrow \nu)}{\mc{E}_n(\bb 0)}
\nn
&= -Q_{\mu} Q_{\nu} \limu \sum_{n=2}^{\norb} \frac{\bra{z_1(\bb 0)}\dm V(\bb 0) \ket{z_n(\bb 0)}\bra{z_n(\bb 0)}\dn V(\bb 0) \ket{z_1(\bb 0)}}{\mc{E}_n(\bb 0)}
\nn
&= 0,
\ea
where the higher-order terms have been omitted.
Defining $\ket{v} = Q_{\mu} \dm V(\bb 0) \ket{z_1(\bb 0)}$, the desired condition reduces to
\ba
\limu \Esec
= -\limu \sum_{n=2}^{\norb} \frac{\brk{v}{z_n(\bb 0)} \brk{z_n(\bb 0)}{v}}{\mc{E}_n(\bb 0)}
= 0.
\label{seq:cond1}
\ea
Note that $\kbr{z_n(\bb 0)}{z_n(\bb 0)}$ is a positive semidefinite matrix.
Thus, \eq{seq:psdnull} implies that \eq{seq:cond1} is equivalent to $\limu \ket{z_n(\bb 0)} \brk{z_n(\bb 0)}{v} = 0$ for any $\bQ$ and $1 < n \leq \norb$.
Since this condition also holds for $n = 1$, we find that
\ba
\limu \Dsec = 0
\quad \lrar \quad
P^0 \dm V(\bb 0) \ket{z_1(\bb 0)} = 0 \quad (^{\forall} \mu),
\ea
where $P^0$ is the projector onto the null space of the $U$-dependent part of $\Hse(\bb 0)$.
Unfortunately, we are not aware of a simple solution to this problem, and believe that it requires exactly solving the $\norb - 1$ gapped magnon modes at $\bQ = \bb 0$ for $U \rightarrow \infty$.
However, model calculations suggest that the spin stiffness saturates to $\Dpert$ if the \textit{minimal metric condition} is satisfied~\cite{huhtinen2022revisiting,herzog2022many,jonah}.
To understand this concept, note that the spin excitation spectrum calculated from \eq{seq:Heff} is invariant upon transformations which preserve the hopping amplitudes while shifting the orbital positions.
Physically, this is because the many-body spectrum of $H$ is invariant under such transformations.
However, this transformation affects the quantum metric and the maximally localized Wannier functions, which in turn impacts $\Epert$.
Thus, one cannot directly equate the quantum metric to physical quantities independent of orbital positions.
Instead, one must use the precise orbital positions which minimizes the quantum metric.
Note that the minimal condition is satisfied if the orbitals originate from maximal Wyckoff positions including rotation (2D) or inversion symmetries.
This is precisely the case in the kagom\'{e} lattice model and the flat band model in the main text, where the orbitals are placed on 3-fold rotation symmetric points.
%

\begin{figure}[t]
	\centering\includegraphics[width=.9\textwidth]{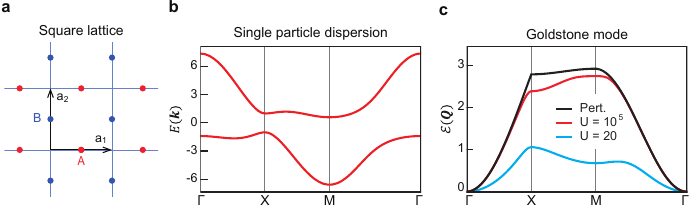}
	\caption{{\bf Spin excitations in the square lattice model.}
		(a) The square lattice in wallpaper group $p4$.
		$s$ orbitals are placed on the $2c$ maximal Wyckoff positions, which guarantees the minimal metric condition.
		(b) Single-particle electron dispersion of the square lattice model (\eq{seq:hsqmin}) with $t_1 = -1.2-0.7i, t_2 = -0.5+2.6i, v_1 = -1.7-0.4i, v_2 = -1.6 + 0.9i$.
		(c) Goldstone mode of the magnon spectrum for $\nocc = 1$ occupied band at different interaction strengths.
		At large $U$ (red), the Goldstone mode saturates to the first order perturbation (black) around $\bQ = \bb 0$.
	}
	\label{sfig1}
\end{figure}

\begin{figure}[h]
	\centering\includegraphics[width=.9\textwidth]{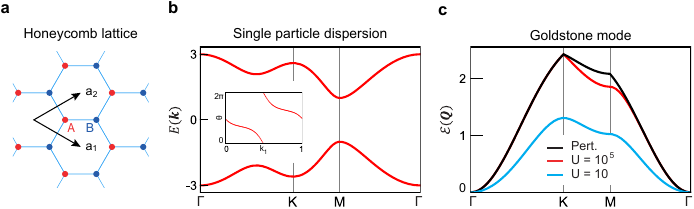}
	\caption{{\bf Spin excitations in the Haldane model.}
		(a) The honeycomb lattice.
		The $A$ and $B$ sites are placed on three-fold rotation invariant points, which enforces the minimal metric condition.
		(b) Single-particle electron dispersion of the Haldane model (\eq{seq:haldane}) with $t_1 = 1, t_2 = 0.5$.
		The inset is the Wilson loop eigenvalue plotted along the $k_1$ direction (see \eq{seq:momentum}), which shows that the occupied band has Chern number $-1$.
		(c) Goldstone mode of the magnon spectrum for $\nocc = 1$ occupied band.
		At large $U$ (red), the Goldstone mode saturates to the first order perturbation (black) around $\bQ = \bb 0$.
	}
	\label{sfig2}
\end{figure}

%
As a demonstration, let us consider the wallpaper group $p4$, and place $s$ orbitals on the $2c$ maximal Wyckoff position as in \fig{sfig1}(a).
The onsite two-fold rotation symmetries enforce the minimal metric condition.
Up to next-nearest neighbor hoppings, the single-particle Hamiltonian is given by
\ba
h(\bk) =2
\bpm
t_1 \cos 2\pi k_1 + t_2 \cos 2\pi k_2 & (t_3-i t_4)\cos \pi(k_1-k_2) + (t_3 + i t_4) \cos \pi(k_1 + k_2) 
\\
(t_3 + i t_4)\cos \pi(k_1-k_2) + (t_3 - i t_4) \cos \pi(k_1 + k_2) & t_1 \cos 2\pi k_2 + t_2 \cos 2\pi k_1
\epm
\label{seq:hsqmin}
,
\ea
which is the most general form containing up to next-nearest neighbor (NNN) hopping amplitudes.
We plot the electron dispersion of $h(\bk)$ in \fig{sfig2}(b).
The Goldstone mode dispersion in \fig{sfig1}(c) shows the saturation of the upper bound at $U \rightarrow \infty$.

As another example, consider the Haldane model~\cite{haldane1988model} in \fig{sfig2}(a), which obeys the minimal metric condition due to the onsite three-fold symmetry.
The single-particle Hamiltonian is characterized by
\ba
h(\bk) =
t_1
\bpm
0 & f(\bk)\\
f^*(\bk) & 0
\epm
+
2 t_2 (-\sin 2\pi k_1 + \sin 2 \pi k_2 + \sin 2\pi (k_1 - k_2)) \sg_z,
\label{seq:haldane}
\ea
where $f(\bk) = e^{-\frac{2\pi i}{3} (2k_1 - k_2)} + e^{\frac{2\pi i}{3} (k_1 + k_2)} + e^{\frac{2\pi i}{3} (k_1 - 2k_2)}$, where $t_1$ is the nearest-neighbor (NN) hopping amplitudes and $t_2$ represents the NNN hopping which breaks the time reversal symmetry.
In \fig{sfig2}(b), we show the single particle band structure for $t_1 = 1$ and $t_2 = 0.5$.
In the inset, we plot the Wilson loop eigenvalue $\theta(k_1)$ obtained by the path-ordered integral of the Berry connection along the $\bb b_2$ direction (see \eq{seq:momentum}), which shows that the occupied band has Chern number $-1$~\cite{bradlyn2022lecture}.
Due to the Wannier obstruction imposed by the nontrivial topology, this system has a large quantum metric integral, which stabilizes the Goldstone mode of the spin excitation in the strongly correlated limit (\fig{sfig2}(c)).
A comparison between the $\Epert$ (black) and the true Goldstone mode energy at $U = 10^5$ (red) clearly shows the validity of the approximation scheme in the strongly correlated limit.

\phantom{}

\section{Numerical calculations on the no-go theorem}
\label{app:nogo}

\begin{figure}[h]
	\centering\includegraphics[width=.5\textwidth]{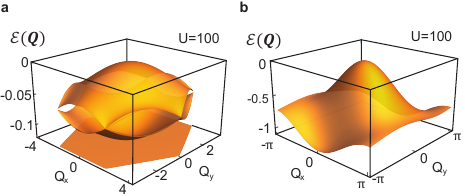}
	\caption{{\bf Negative spin-wave energy of half-filled Hubbard models.}
		(a) The kagom\'{e} lattice with $t = 1$ and $U=100$. 
		(b) The square lattice model introduced in \eq{seq:hsq} with $t_1 = -1.2-0.7i, t_2 = -0.5+2.6i, v_1 = -1.7-0.4i, v_2 = -1.6 + 0.9i$, and $U=100$.
	}
	\label{sfig3}
\end{figure}

In this Appendix, we demonstrate the validity of the no-go theorem by calculating the magnon dispersion of half-filled models.
\fig{sfig3}(a) describes the magnon spectrum of the NN kagom\'{e} lattice at half filling.
The magnon bands with negative energy indicate the instability of saturated ferromagnetism.
\fig{sfig3}(b) illustrates a representative example of a half-filled system with a single orbital, which is obtained from a 2D square lattice.
To show the generality of the no-go theorem, we introduce arbitrary complex hopping amplitudes up to NNN, and break every symmetry other than SU(2) symmetry.
The noninteracting Hamiltonian is given by
\ba
h(\bk) = t_1 \ee{i k_x} + t_2 \ee{i k_y}+ v_1 \ee{i(k_x-k_y)} + v_2\ee{i(k_x + k_y)} + c.c.,
\label{seq:hsq}
\ea
where $t_i$ and $v_i$ denote the NN and NNN hopping amplitudes, respectively.
The magnon spectrum with negative energy shown in \fig{sfig3}(b) clearly indicates that saturated ferromagnetism is prohibited.
%

\section{Mean-field Hamiltonian of saturated ferromagnets}
\label{app:Hmf}
In this Appendix, we briefly mention the mean-field Hamiltonian used for calculating the spin-resolved electron band dispersion in \fig{fig1}(b).
Following the main text, we assume that every occupied state has spin $\ua$.
The mean-field Hamiltonian obtained by decoupling the many-body terms of the second line of \eq{seq:Hubbard} is given by
\ba
H_{mf}=\sum_{\bk \al \be \sg} h(\bk)_{\al \be} \cdag_{\bk \al \sg} c_{\bk \be \sg}
+
U \sum_{\bk \al \sg} \expv{n_{\al -\sg}} \cdag_{\bk \al \sg} c_{\bk \al \sg} 
-
U \nc \sum_{\al} \expv{n_{\al \ua}} \expv{n_{\al \da}},
\label{seq:Hmf}
\ea
where $\expv{n_{\al \sg}} = \expv{\cdag_{i \al \sg} c_{i \al \sg}}$ is the unit cell-independent orbital filling to be self-consistently determined.
In the $(\ua, \da)^T \otimes (\al_1, \dots, \al_{\norb})^T$ basis, $H_{mf} (\bk)$ is a $(2 \norb \times 2 \norb)$ Hermitian matrix given by
\ba
H_{mf}(\bk)
=
\bpm
h_{\ua \ua} (\bk) & h_{\ua \da} (\bk) \\
h_{\da \ua} (\bk) & h_{\da \da} (\bk) \\
\epm
=
\bpm
h(\bk) & 0 \\
0 & h(\bk) \\
\epm
+
U
\bpm
diag (\bb{n}_{\da}) & 0 \\
0 & diag (\bb{n}_{\ua}) \\
\epm
,
\label{seq:Hmfk}
\ea
where $\bb{n}_{\sg} = (\expv{n_{1 \sg}}, \dots, \expv{n_{\norb \sg}})$.
Since the $\da$ bands are empty ($\bb{n}_{\da}=\bb{0}$), the Hamiltonian of occupied states (which have $\ua$) is identical to $h(\bk)$.
This observation underscores that the ground state of a saturated ferromagnet is fully determined by the noninteracting Hamiltonian.
As an example, we present the mean-field band structure of a kagom\'{e} ferromagnet in \fig{fig1}(b) of the main text.

\end{document}